\documentclass[aps,prb,a4paper,twocolumn,showpacs,floatfix,citeautoscript]{revtex4-1}
\usepackage{graphicx}
\usepackage{graphicx}
\usepackage{amsfonts}
\usepackage{amsmath}
\usepackage{amssymb}
\usepackage{bm} 
\usepackage{bbold}
\usepackage{mathtools}
\usepackage{dcolumn}
\usepackage{hyperref}
\usepackage[usenames]{color}

\usepackage{subcaption}

\usepackage{ragged2e}
\DeclareCaptionJustification{justified}{\justifying}
\newcommand{\pr}{%
        ^\prime}

\newcommand{\svek}{%
        \mathbf}

\newcommand{\out}[1]{{}}

\newcommand{\fref}[1]{Fig.~\ref{#1}}
\newcommand{\sref}[1]{Sec.~\ref{#1}}

\newcommand{\idmatrix}{{\mathbf{1}}}

\newcommand{\op}[1]{\widehat{#1}}
\newcommand{\kvec}[1]{{\mathbf{#1}}}
\newcommand{\cvec}[1]{{\mathrm{#1}}}
\newcommand{\tsymb}{{\mathcal{T}}}
\newcommand{\torder}[1]{\left\langle \! \tsymb \! \left[#1\right] \! \right\rangle}

\begin{document}

\title{{\em Ab initio} dynamical vertex approximation}
\author{Anna Galler, Patrik Thunstr\"om, Patrik Gunacker, Jan M. Tomczak, Karsten Held}
\affiliation{Institute of Solid State Physics,  TU Wien, A-1040 Vienna, Austria}

\begin{abstract} 
Diagrammatic extensions of dynamical mean field theory (DMFT) such as the dynamical vertex approximation (D$\Gamma$A) allow us to include non-local correlations beyond DMFT on all length scales and proved their worth 
for model calculations. Here, we develop  and implement an AbinitioD$\Gamma$A approach for electronic structure calculations of materials. Starting point is the two-particle irreducible vertex in the two particle-hole channels which is approximated by the bare non-local Coulomb interaction and all local vertex corrections.  From this we calculate the full non-local vertex and the non-local self-energy through the Bethe-Salpeter equation.  The  AbinitioD$\Gamma$A approach naturally generates all local DMFT correlations and all non-local $GW$ contributions, but also further non-local correlations beyond: mixed terms of the former two and non-local spin fluctuations.
We apply this new methodology to the prototypical correlated metal SrVO$_3$.
\end{abstract}

\maketitle

\section{Introduction}

Some of the most fascinating physical phenomena are experimentally
observed in strongly correlated electron systems and, on the theoretical side,
only poorly understood hitherto. This is particularly true for electronic structure calculations of materials where the standard approach, density functional theory (DFT) \cite{Hohenberg1964,Kohn1965,Jones1989a,Martin04} in its local density approximation
(LDA) or generalized gradient approximation (GGA)
only rudimentarily includes such correlations. This calls for genuine many body techniques \cite{Martin16}.

\begin{figure}
\includegraphics[clip=true,width=8.5cm]{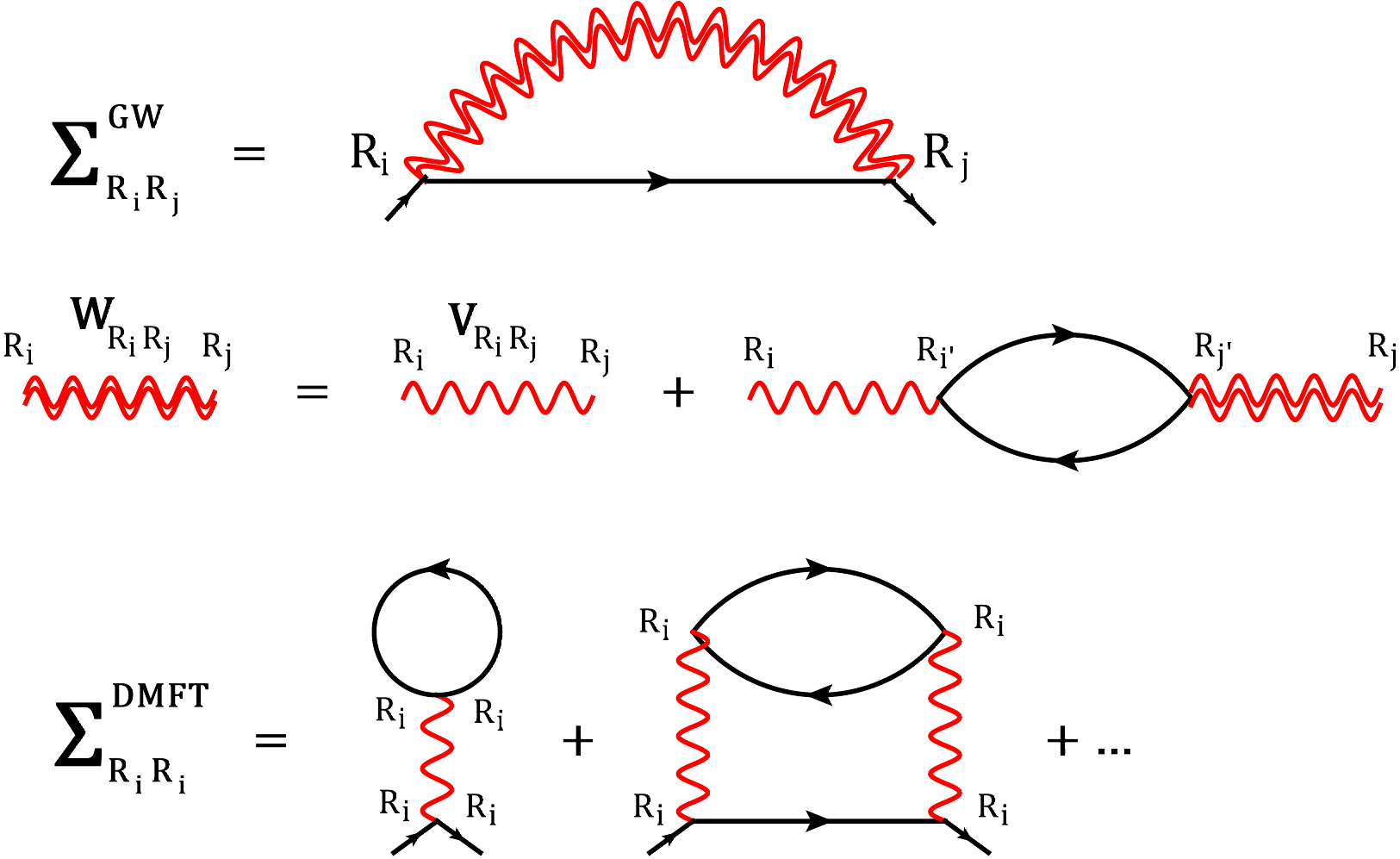}

\caption{(Color online) {Top:}
In addition to the  Hartree term,  $GW$  takes into
account the (screened) exchange Feynman diagram (wiggled line: Coulomb interaction $V$; double wiggled line: screened interaction $W$;  line: interacting Green's function $G$;  $\svek{R}_i$, $\svek{R}_j$ indicate the lattice sites).
{Middle:} Calculation of $W$ as a screened  Coulomb interaction within the RPA. {Bottom:} Feynman diagrams for the DMFT self-energy $\Sigma$.
\label{Fig:GWDMFT}}
\end{figure}

One such method is Hedin's   $GW$ approach \cite{Hedin1965} 
consisting of the interacting Green's function times the screened interaction $W$,
which physically  describes a screened exchange, see  Fig.\ \ref{Fig:GWDMFT} top panel. 
In the last years, this approach
has matured to the point that material calculations are actually feasible and
various program packages are available. As a consequence, semiconductors,
in which the extended $sp^3$ orbitals make the non-local exchange contribution
particularly important, can be better described, especially their band gaps.
From the point of view of 
the exchange-correlation potential of DFT,
the $GW$ approach mostly improves upon the LDA or GGA regarding the exchange part.
Via the inclusion of screening, $GW$ implicitly also includes correlation effects, leading to renormalized quasi-particle weights and finite life times.
Nonetheless, in the presence of strong electronic correlations, e.g.\ in transition metal oxides and $f$-electron systems, 
the first order expression $GW$ of many-body perturbation theory is largely insufficient and vertex corrections become relevant.

For such strongly correlated materials, dynamical mean field theory (DMFT) \cite{Metzner1989,Georges1992a,Georges1996} emerged  instead as the state-of-the-art.
 The reason for
this is that DMFT accounts for a major part of the electronic correlations, namely the local correlations between electrons on the same lattice site. These are particularly strong for transition metal oxides or heavy fermion systems with $d$- and $f$-electrons, respectively, due to
the localized nature of the corresponding orbitals. 
Its merger with LDA \cite{Anisimov1997,Lichtenstein1998,Kotliar2006,Held2007} or $GW$ \cite{Biermann2003,Tomczak2012,Taranto2013,PhysRevB.88.235110,Tomczak14,jmt_sces14,Choi2016,Boehnke2016,Roekeghem2014} allows for realistic materials calculations and is more and more widely used. Does the principal method development of electronic structure calculations come  to a standstill at this point? Or does it merely advance towards ever more complex and bigger systems?

In this paper, we show that a further big step forward
is possible. Let us, to this end, start by analyzing $GW$ and DMFT, which are both based on Feynman diagrams: $GW$ simply takes (besides the Hartree term) the exchange diagram (Fig.\ \ref{Fig:GWDMFT} top) and much of its strengths result from the fact that this exchange term is taken in terms of the screened Coulomb interaction within the random phase approximation (RPA; Fig.\ \ref{Fig:GWDMFT} middle). This screening results in a much better convergence of the perturbation series  of which actually only the first order terms are taken into account. DMFT, on the other hand, includes all local (skeleton) diagrams for the self-energy in terms of the interacting local Green's function and Hubbard/Hund-like local interactions (Fig.\ \ref{Fig:GWDMFT} bottom). While this reliably accounts for the local electronic correlations, non-local correlations are neglected in DMFT. 
The same holds for extended DMFT \cite{Si1996,Chitra00} which treats the local correlations emerging from non-local interactions.
The non-local correlations are, however, at the heart of some of the most fascinating phenomena associated with electronic correlations
such as (quantum) criticality, spin fluctuations and, possibly, high-temperature superconductivity. 

In this paper, we  develop, implement and apply a 21$^{\hbox{\tiny st}}$ century method for the {\sl ab initio} calculation of correlated materials.
It is based on recent diagrammatic extensions of DMFT \cite{Kusunose06,Toschi07,Katanin2009,Rubtsov2008,Slezak2009,Rohringer2013,Taranto2014,Kitatani2015,Ayral2015,Li2015,Valli2015,Li2016}, a development which started with the dynamical vertex approximation (D$\Gamma$A) \cite{Toschi07,Katanin2009}. These dynamical vertex approaches are quite similar and all based on the two-particle vertex instead of the one-particle vertex  (i.e.\ the self-energy) in DMFT.
This way, local dynamical correlations {\`a} la DMFT are captured 
but at the same time strong electronic correlations on all time and length scales are also included.
In the context of many-body models, D$\Gamma$A and related approaches have been applied successfully to calculate, among others,
(quantum) critical exponents,\cite{Rohringer2011,Antipov2014,Hirschmeier2015,Schaefer2016} and evidenced strong non-local contributions to the self-energy beyond {\it GW}.\cite{Schaefer2015}

\begin{figure*}[t!]
\begin{minipage}{18 cm}
\includegraphics[clip=true,trim=60 40 40 20,width=18.cm]{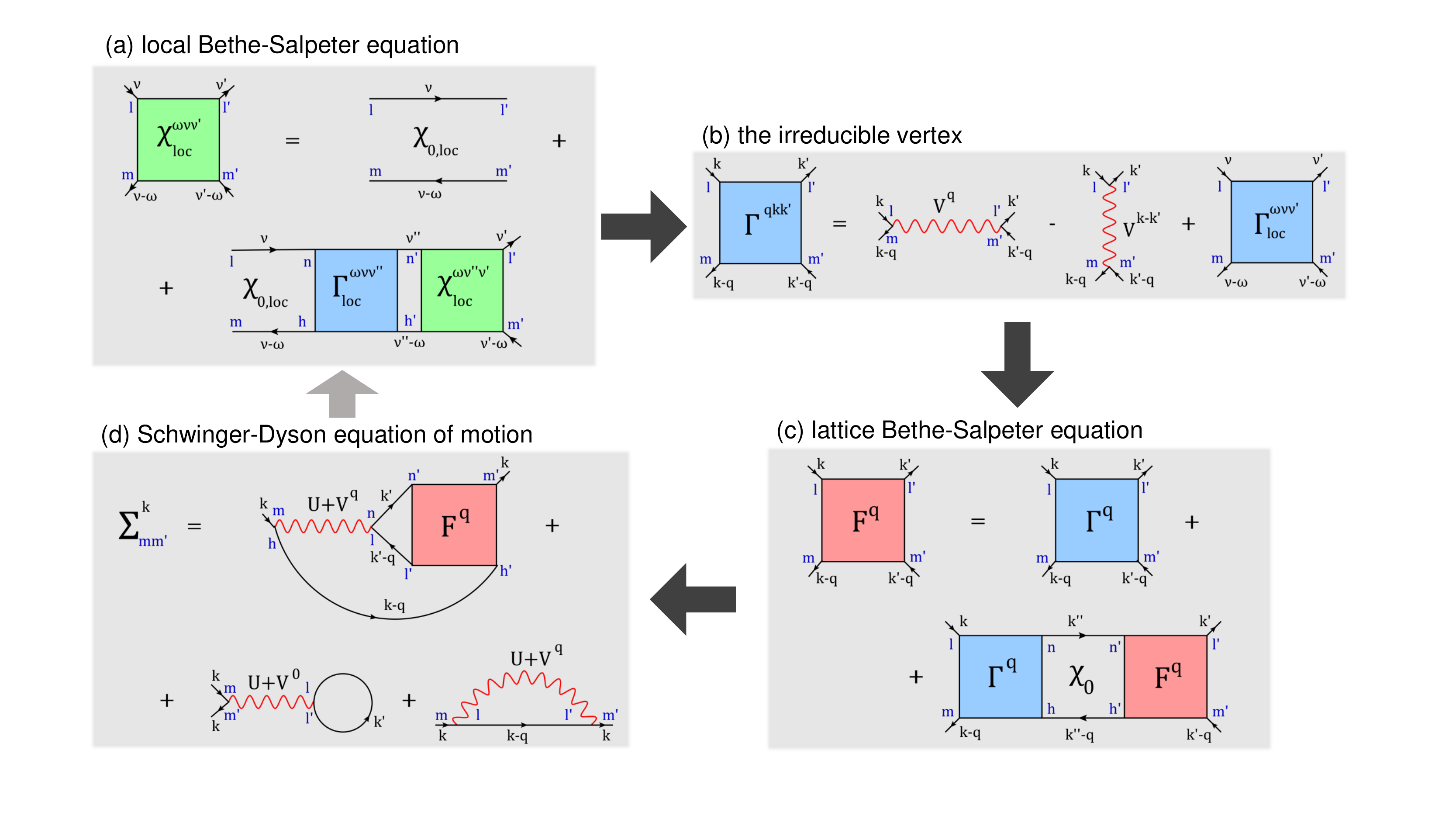}
\caption{(Color online) Outline of the main equations of the AbinitioD$\Gamma$A approach: 
(a) The local Bethe-Salpeter equation allows extracting
the local irreducible vertex $\Gamma_{loc}$ from the local generalized susceptibility $\chi_{loc}$.
 (b) The local irreducible vertex (which contains the local interaction $U$)
is supplemented by the non-local interaction $V$ resulting in the momentum-dependent  irreducible vertex $\Gamma$. 
(c) From this $\Gamma$ the full vertex $F$ is obtained using the Bethe-Salpeter equation (shown is only the particle-hole channel, but the related
contribution from the transversal particle-hole channel is also included). 
(d) Finally, the self-energy is constructed via the Schwinger-Dyson equation consisting of
the vertex part (top), and the Hartree-Fock  contribution (bottom).
From this self-energy one can, in principle, determine an updated local vertex,
closing the loop.
 For details and the index convention, see \sref{diagext} and \sref{EoM}.
\label{Fig:flow}}
\end{minipage}
\end{figure*}

One can also consider the first principles extension AbinitioD$\Gamma$A as a realization of Hedin's  idea  \cite{Hedin1965} to include vertex corrections beyond the $GW$ approximation. All vertex corrections which can be traced back to the irreducible local vertex in the particle-hole channels and the bare non-local Coulomb interaction are included, see Fig. \ref{Fig:flow}(b). This seamlessly generates all the $GW$ diagrams 
and the associated physics, as well as the local diagrams of DMFT and non-local
correlations beyond both  on all length scales. Through the latter, we can 
 describe, among others, phenomena such as quantum criticality, spin-fluctuation mediated superconductivity, and weak localization corrections to the conductivity. This is beyond DMFT which is restricted to local correlations as well as beyond $GW$ which is restricted to one screening channel and the low-coupling regime.\cite{Notevertexcorr}
Nonetheless, the computational effort of AbinitioD$\Gamma$A is still manageable even for materials calculations with several relevant orbitals, as we demonstrate in this work.

In Section \ref{Sec:method}, we introduce the AbinitioD$\Gamma$A method, including all relevant equations. Section  \ref{Sec:results}, presents first results for the testbed material SrVO$_3$; and  Section \ref{Sec:outlook} summarizes the work and provides an outlook for future applications. 	An avenue to 
 AbinitioD$\Gamma$A was envisioned in Ref.\ \onlinecite{Toschi2011}. Here we concretize these ideas and fully derive and implement the approach. Please also note the proposal of Ref.~\onlinecite{Katanin2016}
 to use the functional renormalization group on top of the (extended) DMFT, and the dual boson approach \cite{Rubtsov12} to non-local interactions.

\section{AbinitioD$\Gamma$A method} 
\label{Sec:method}

Before we go into the detailed multi-orbital derivation of the AbinitioD$\Gamma$A equations, 
let us briefly outline the rationale of the method, as it is depicted in \fref{Fig:flow}.

As a starting point we consider 
a general Hamiltonian written in terms of a  one-particle operator ($\op{H}_0$), and a two-particle interaction with a local ($\op{U}$) as well as a  non-local ($\op{V}^\kvec{q}$) part:
\begin{equation}
\op{H} = \op{H}_0 + \op{U} + \sum_{\kvec{q}}\op{V}^{\kvec{q}}.
\end{equation}
For  AbinitioD$\Gamma$A calculations this Hamiltonian may contain a large set of orbitals, e.g., the physical $spdf$ orbitals in a muffin tin orbital  basis set \cite{Andersen2000} or those obtained by a Wannier function projection \cite {Kuneifmmodecheckselsevsfi2010a,Mostofi2008}.
However, the approach can also be applied to a more restricted set of orbitals for the low energy degrees of freedom such as the $t_{2g}$ orbitals for our SrVO$_3$ calculation  in Section \ref{Sec:results}. 
In the latter case, the influence of orbitals outside the energy windows of $\op{H}_0$ and their effect on $\op{H}_0$ have to be taken into account, e.g.,
on the DFT or $GW$ level. The screening of the interactions $\op{U}$ and $\op{V}^{\kvec{q}}$  needs to be included as well, e.g., through constrained DFT \cite{Dederichs84,McMahan88,Gunnarsson89} or constrained RPA \cite{PhysRevB.70.195104, miyake2008screened, mno}.

AbinitioD$\Gamma$A is a Feynman diagrammatic theory built around the two-particle irreducible vertex which is  approximated by the bare Coulomb interaction plus all local vertex corrections, see Fig. \ref{Fig:flow}(b).
From this irreducible vertex many additional Feynman diagrams are constructed. One has to distinguish between (i) the fully irreducible vertex as a starting point where these additional diagrams are constructed by the parquet equations as in Refs.\ \onlinecite{Valli2015,Li2016} and (ii) the irreducible vertex in the particle-hole channel (and transversal particle hole channel) where this is done by the Bethe-Salpeter equations (BSE) as in Ref.\ \onlinecite{Toschi07,Katanin2009}. For our realistic multi-orbital calculations we here rely on (ii) which is numerically more feasible.

To include the important local vertex corrections, first the local irreducible vertex $\Gamma_{loc}$ is extracted from the local two-particle Green's function 
or from generalized susceptibilities $\chi_{loc}$
(see \fref{Fig:flow}(a)). 
This is possible by solving an  Anderson impurity model; and for multi-orbitals, continuous-time quantum Monte Carlo simulations \cite{Gull2011a,Boehnke14,Gunacker15,Gunacker16} are most appropriate to this end. The resulting local irreducible vertices in the longitudinal and transversal particle-hole channels are then combined with the non-local interaction $V^q$ and finally dressed via the BSE equation (see Fig.\ \ref{Fig:flow}(c);  \sref{diagext}).
Eventually, a new lattice Green's function is constructed using the $\svek{k}$-dependent self-energy calculated from the equation of motion (see Fig.\  \ref{Fig:flow}(d);  \sref{EoM}).

In principle from the local projection of this new Green's function an updated local vertex can be calculated as indicated by the arrow from  Fig.\  \ref{Fig:flow} (d) to  Fig.\  \ref{Fig:flow} (a). Such a self-consistent scheme has been envisaged in Refs.\ \onlinecite{Held2008,Held2014,Ayral2016} but not yet implemented; it is of particular importance if the electron density changes considerably in D$\Gamma$A because the vertex and its asympotics depend strongly on the density. Beyond this, also the  DFT Hamiltonian  or the constrained RPA interaction and $GW$ self-energy  for the high energy degrees of freedom should be updated through a charge \cite{Savrasov04,Minar05,Pouroskii07,Aichhorn2011,Bhandary16} or hermitianized self-energy \cite{P7:Faleev04,P7:Chantis06}
 self-consistency, respectively.  These steps go beyond the one-shot calculation of the present paper where  the local vertex is fixed to the  DFT+DMFT solution.

As the computationally most demanding part is the calculation of the local vertex,
it is reasonable (in case of a large set of orbitals)  to calculate it only for the more correlated (e.g., $d$ and $f$) orbitals, whereas the local vertex of the less correlated (e.g.,  $spd$) orbitals may be taken as  $U+V$
in the same way as the two $V$ terms  in Fig.\ \ref{Fig:flow} (b).  This includes all the {\it GW} diagrams for these orbitals but also Feynman diagrams beyond.
\cite{NotebeyondGW}
 A frequency dependence of $U(\omega)$ when using the constrained RPA as a starting point can also be included for the more correlated orbitals in the 
same way, i.e., adding $U(\omega)-U$ to the vertex.\cite{noteHirayama} Alternatively, one can calculate the local vertex from $U(\omega)$ in CT-QMC.

Let us after these general considerations, now turn to the actual equations and technical details of the AbinitioD$\Gamma$A approach.
Fig.\ \ref{Fig:flow} provides an overview, but the devil is in the details
and Fig.\ \ref{Fig:flow}  is somewhat schematic: We have not specified the spin-indices and have only shown the longitudinal (not the transversal) particle-hole channel; also, in our implementation, we
circumvent an explicit evaluation of $\Gamma_{loc}$ as this quantity may contain divergencies; we also show how to increase the numerical efficiency by a reformulation in terms of three-leg quantities and by neglecting---as an additional approximation---the $\kvec{k},\kvec{k}'$ dependence of the irreducible vertex in Fig.\ \ref{Fig:flow} (b).

\subsection{Coulomb interaction}
The electron-electron Coulomb interaction $\op{U}^{{\mathrm{full}}}$ can in general be expressed as
\begin{eqnarray}
\op{U}^{{\mathrm{full}}} &=& \frac{1}{2}\sum_{\substack{ \svek{R}_1,\svek{R}_2,\svek{R}_3 \\ ll\pr mm\pr \\ \sigma \sigma' }} U^{{\mathrm{full}}}_{lm\pr ml\pr}(\svek{R}_1,\svek{R}_2,\svek{R}_3)\times\\
 &\times& \op{c}^\dagger_{{\svek{R}_3m\pr \sigma}}  \op{c}^\dagger_{\svek{R}_1l\sigma'} \op{c}^{\phantom{\dagger}}_{\svek{R}_2m\sigma'} \op{c}^{\phantom{\dagger}}_{\svek{0}l\pr\sigma}, \nonumber
\end{eqnarray}
where the Roman indices $ll\pr mm\pr$ denote the orbitals, $\sigma$  the spin, and $\svek{R}$ the lattice site. It fulfills the particle ``swapping symmetry''
\begin{equation}
U^{{\mathrm{full}}}_{lm\pr ml\pr}(\svek{R}_1,\svek{R}_2,\svek{R}_3) = U^{{\mathrm{full}}}_{m\pr ll\pr m}(\svek{R}_3-\svek{R}_2,-\svek{R}_2,\svek{R}_1-\svek{R}_2),
\end{equation}
which corresponds to an invariance under a swap of both the incoming and the outgoing particle labels. Taking the Fourier transform with respect to $\svek{R}$ yields
\begin{align}
U^{\kvec{q}\kvec{k}\kvec{k}'}_{lm\pr ml\pr} & = \sum_{\svek{R}_1,\svek{R}_2,\svek{R}_3} e^{i\kvec{k}\svek{R}_1}e^{-i(\kvec{k}-\kvec{q})\svek{R}_2}e^{i(\kvec{k}\pr - \kvec{q})\svek{R}_3}\nonumber \\ &  \times U^{{\mathrm{full}}}_{lm\pr ml\pr}(\svek{R}_1,\svek{R}_2,\svek{R}_3) \; ;
\end{align}
or for the interaction operator
\begin{equation}
\op{U}^{{\mathrm{full}}} = \frac{1}{2}\sum_{\substack{ \kvec{q}\kvec{k}\kvec{k}'\\ ll\pr mm\pr \\ \sigma \sigma' }}  U^{\kvec{q}\kvec{k}\kvec{k}'}_{lm\pr ml\pr} \op{c}^\dagger_{\kvec{k}'-\kvec{q}m\pr \sigma}  \op{c}^\dagger_{\kvec{k} l\sigma'} \op{c}^{\phantom{\dagger}}_{\kvec{k}-\kvec{q}  m \sigma'} \op{c}^{\phantom{\dagger}}_{ \kvec{k}' l\pr \sigma},
\end{equation}
where
\begin{align}
\op{c}_{\kvec{k} l \sigma } & = \sum_{\svek{R}} e^{i\kvec{k}\svek{R}} \op{c}_{\svek{R} l \sigma}.
\end{align}

The k-point dependence of $\op{U}^{{\mathrm{full}}}$ can be simplified if the orbital overlap between adjacent unit-cells is neglected, so that the creation and annihilation operators are paired up at site $\svek{0}$ and $\svek{R}$. This gives
\begin{align}
U_{lm\pr ml\pr} & \equiv U^{{\mathrm{full}}}_{lm\pr ml\pr}(\svek{0},\svek{0},\svek{0}),\\
V^{\kvec{q}}_{lm\pr ml\pr} & \equiv \sum_{\svek{R} \neq 0} e^{i\svek{R}\kvec{q}} U^{{\mathrm{full}}}_{lm\pr ml\pr}(\svek{R},\svek{R},\svek{0}),
\end{align}
which corresponds to a local interaction $\op{U}$ and a purely non-local interaction $\op{V}^\kvec{q}$. In this case the swapping symmetry reduces to $U_{lm\pr ml\pr} = U_{m\pr ll\pr m}$ and $V^{\kvec{q}}_{lm\pr ml\pr} = V^{-\kvec{q}}_{m\pr ll\pr m}$. 

\subsection{Green's functions}
We begin with the basic definitions of the one- and two-particle Green's functions
\begin{align}
G^{\kvec{k}}_{\sigma,lm}(\tau) & \equiv -\torder{ \op{c}^{\phantom\dagger}_{\svek{k} l\sigma}(\tau) \op{c}^{\dagger}_{\svek{k} m\sigma }(0) },\\
G^{\kvec{q}\kvec{k}\kvec{k}'}_{\substack{ lmm\pr l\pr \\ \sigma_1\sigma_2\sigma_3\sigma_4 }}(\tau_1,\tau_2,\tau_3) & \equiv  \\
\span\span\torder{  \op{c}^{\phantom\dagger}_{\svek{k} l\sigma_1 }(\tau_1) \op{c}^{\dagger}_{\kvec{k} - \kvec{q}  m\sigma_2 }(\tau_2)  \op{c}^{\phantom\dagger}_{\kvec{k}'-\kvec{q}  m\pr\sigma_3 }(\tau_3) \op{c}^{\dagger}_{\svek{k}' l\pr\sigma_4}(0) }\nonumber.
\end{align}
where $\tau \in [0,\beta)$ denotes imaginary time and $\tsymb$ is the time ordering operator. In absence of spin-orbit interaction the spin is conserved, which leaves 6 different spin combinations
\begin{align}
G^{\kvec{q}\kvec{k}\kvec{k}'}_{\sigma\sigma',lmm\pr l\pr}(\tau_1,\tau_2,\tau_3) & \equiv G^{\kvec{q}\kvec{k}\kvec{k}'}_{\substack{ lmm\pr l\pr \\ \sigma\sigma \sigma'\sigma' }}(\tau_1,\tau_2,\tau_3),\\
G^{\kvec{q}\kvec{k}\kvec{k}'}_{\overline{\sigma\sigma'},lmm\pr l\pr}(\tau_1,\tau_2,\tau_3) & \equiv G^{\kvec{q}\kvec{k}\kvec{k}'}_{\substack{ lmm\pr l\pr \\ \sigma\sigma' \sigma'\sigma }}(\tau_1,\tau_2,\tau_3).
\end{align}
There are only two independent spin configurations in the paramagnetic phase, as the system then is SU(2) symmetric with respect to the spin, 
\begin{align}
G_{\sigma\sigma'} & = G_{(-\sigma)(-\sigma')} = G_{\sigma'\sigma},\\
G_{\sigma\sigma} & = G_{\sigma(-\sigma)} + G_{\overline{\sigma(-\sigma)}}\label{eqn:su2}.
\end{align}
As we will see in the next section, one particularly useful choice for these two spin combinations is the density and magnetic channel defined as
\begin{align}
G_{d} & = G_{\uparrow\uparrow} + G_{\uparrow\downarrow}, \label{eqn:d}\\
G_{m} & = G_{\uparrow\uparrow} - G_{\uparrow\downarrow} = G_{\overline{\uparrow\downarrow}}. \label{eqn:m}
\end{align}

The value of the two-particle Green's function takes a step of 1 whenever the $\tau$ arguments of a creation and an annihilation operator become equal. These discontinuities can be cancelled out by subtracting pairs of one-particle Green's functions, giving the so-called connected part of the two-particle Green's function
\begin{eqnarray}
&&G^{{\mathrm{con~}} \kvec{q}\kvec{k}\kvec{k}'}_{\sigma\sigma', lmm\pr l\pr}(\tau_1,\tau_2,\tau_3)  = G^{\kvec{q}\kvec{k}\kvec{k}'}_{\sigma\sigma', lmm\pr l\pr}(\tau_1,\tau_2,\tau_3)\nonumber \\
&&\quad\;\;\; -\, \delta_{\kvec{q}0} G^{\kvec{k}}_{\sigma,lm}(\tau_1 - \tau_2) G^{\kvec{k}'}_{\sigma',m\pr l\pr}(\tau_3) \nonumber\\
&&\quad\;\;\; +\, \delta_{\sigma\sigma'}\delta_{\kvec{k}\kvec{k}'} G^{\kvec{k}}_{\sigma,ll\pr}(\tau_1) G^{\kvec{k}-\kvec{q}}_{\sigma,m\pr m}(\tau_3-\tau_2).\label{eqn:gcon}
\end{eqnarray}
The connected part is continuous in its $\tau$ arguments, but it still shows cusps at equal times. We define the Fourier transformation of $G^{{\mathrm{con~}}}$ with respect to $\tau$ in the same way as for $\op{U}^{{\mathrm{full}}}$ and $\svek{R}$,
\begin{align}
G^{{\mathrm{con~}} \cvec{q}\cvec{k}\cvec{k}'}_{\sigma\sigma',lmm\pr l\pr} & \!= \! \int_0^\beta\!\! \int_0^\beta\!\! \int_0^\beta\!\! d\tau_1d\tau_2d\tau_3  e^{i\nu\tau_1} e^{-i(\nu - \omega)\tau_2} e^{i(\nu'-\omega)\tau_3} \nonumber\\ 
& \times G^{{\mathrm{con~}} \kvec{q}\kvec{k}\kvec{k}'}_{\sigma\sigma', lmm\pr l\pr}(\tau_1,\tau_2,\tau_3),
\end{align}
where the bosonic compound index is $\cvec{q} = ( \omega, \kvec{q} )$ and the fermionic compound index $\cvec{k} = ( \nu,\kvec{k})$. In the chosen frequency and momentum convention the bosonic index ($\cvec{q}$) corresponds to a longitudinal transfer of energy and momentum from one particle-hole pair ($ml$) to the other ($m\pr l\pr$). 

$G^{{\mathrm{con~}}}$ is by definition related to the fully reducible vertex $F$ as
\begin{equation}
G^{{\mathrm{con~}} \cvec{q}\cvec{k}\cvec{k}'}_{r,lmm\pr l\pr} = \sum_{nn\pr hh\pr} \chi^{\cvec{q}\cvec{k}\cvec{k}}_{0,lmhn}F^{\cvec{q}\cvec{k}\cvec{k}'}_{r,nhh\pr n\pr}\chi^{\cvec{q}\cvec{k}'\cvec{k}'}_{0,n\pr h\pr m\pr l\pr}, \label{eqn:gfromf}
\end{equation}
where the bare 2-particle propagator $\chi_0$ is defined as
\begin{equation}
\chi^{\cvec{q}\cvec{k}\cvec{k}'}_{0,lmm\pr l\pr} \equiv -\beta G^{\cvec{k}}_{ll\pr} G^{\cvec{k}'-\cvec{q}}_{mm\pr} \delta_{\cvec{k}\cvec{k}'}. 
\end{equation}
The full vertex $F$ is part of the definition of the Bethe-Salpeter equation (BSE), and will thus be of major importance in the diagrammatic extension outlined in the next section.

\begin{figure}
\includegraphics[width=8cm]{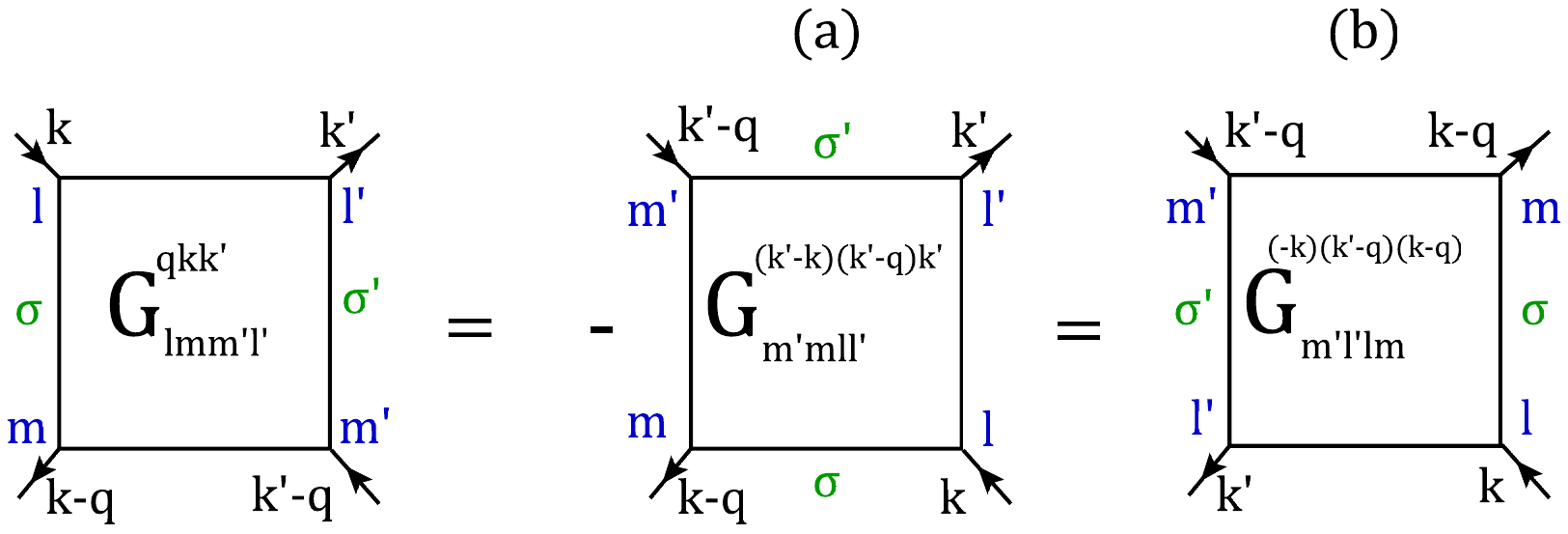}
\caption{Diagrammatic representation of (a) the crossing symmetry in Eq. (\ref{eqn:cross1}) and (b) the swapping symmetry in Eq. (\ref{eqn:swap}). 
\label{fig:crossing}}
\end{figure}

In order to improve the statistics of the two-particle Green's function, and reduce the computational resources needed to perform the calculations, it is important to utilize the symmetries of the system. In addition to the orbital symmetries the two-particle Green's function also fulfills time reversal symmetry,
\begin{equation}
G^{\cvec{q}\cvec{k}\cvec{k}'}_{\sigma\sigma',lmm\pr l\pr} = G^{\bar{\cvec{q}}\bar{\cvec{k}'}\bar{\cvec{k}}}_{\sigma'\sigma,l\pr m\pr m l},
\end{equation}
where $\bar{\cvec{k}} = \{\nu,-\kvec{k}\}$, and the crossing symmetries,
\begin{align}
G^{\cvec{q}\cvec{k}\cvec{k}'}_{\sigma\sigma',lmm\pr l\pr} & = -G^{ (\cvec{k}'-\cvec{k})(\cvec{k}'-\cvec{q})\cvec{k}'}_{\overline{\sigma'\sigma},m\pr m ll\pr} \label{eqn:cross1}\\
& = -G^{(\cvec{k}-\cvec{k}')\cvec{k}(\cvec{k}-\cvec{q})}_{\overline{\sigma\sigma'},ll\pr m\pr m} \label{eqn:cross2} \\
& = G^{(-\cvec{q})(\cvec{k}'-\cvec{q})(\cvec{k}-\cvec{q})}_{\sigma'\sigma,m\pr l\pr lm},\label{eqn:swap}
\end{align}
where the last line corresponds to a full swap of the in-coming and the outgoing particle labels. The symmetries Eqs. (\ref{eqn:cross1}-\ref{eqn:swap}) can be understood  from the fact that exchanging the position of the ``legs'' does not alter the  vertex but the $q,k,k'$ values it corresponds tp as visualized in Fig.\ \ref{fig:crossing}. Finally, the two-particle Green's function transforms under complex conjugation as
\begin{equation}
(G^{\cvec{q}\cvec{k}\cvec{k}'}_{\sigma\sigma',lmm\pr l\pr})^* = G^{(-\cvec{q})(-\cvec{k})(-\cvec{k}')}_{\sigma'\sigma,l\pr m\pr ml}.
\end{equation}

\subsection{Diagrammatic extension}
\label{diagext}
At the heart of the AbinitioD$\Gamma$A method is the two-particle irreducible vertex $\Gamma$ in the particle-hole channel
\begin{align}
\Gamma^{\cvec{q}\cvec{k}\cvec{k}'}_{\sigma\sigma',lmm\pr l\pr} & \equiv \Gamma^{\omega\nu\nu'}_{\sigma\sigma',lmm\pr l\pr} + {\mathbf{V}}^{\kvec{q}\kvec{k}\kvec{k}'}_{\sigma\sigma',lmm\pr l\pr},\label{eqn:gammastart} \\
{\mathbf{V}}^{\kvec{q}\kvec{k}\kvec{k}'}_{\sigma\sigma',lmm\pr l\pr} & \equiv \beta^{-2} (V^{\kvec{q}}_{lm\pr ml\pr} - \delta_{\sigma\sigma'} V^{\kvec{k}'-\kvec{k}}_{mm\pr ll\pr}),\label{eqn:vqkk}
\end{align} 
given by the local irreducible vertex $\Gamma^{\omega\nu\nu'}$ supplemented with the non-local interaction ${\mathbf{V}}^{\kvec{q}\kvec{k}\kvec{k}'}$ written in the form of a fully irreducible vertex, as shown in Fig.\ \ref{Fig:flow}(b).
 For brevity we omit here and in the following
 in $\Gamma^{\omega\nu\nu'}$ a
 ``loc'' subscript [which is implied if the vertex depends on frequencies only; please recall the convention $\cvec{q}=(\omega,\kvec{q})$ and $\cvec{k}=(\nu,\kvec{k})$] and a ``ph'' subscript ($\Gamma$'s and later $\phi$'s without an explicit subscript refer to the particle-hole channel). As already mentioned, $\Gamma$ can be extracted from the solution of an effective Anderson impurity problem through the inversion of a local BSE, which relates the local two-particle irreducible ($\Gamma$) and reducible ($\phi$) vertices in the particle-hole channel with the local full vertex
\begin{align}
F^{\omega\nu\nu'}_{r,lmm\pr l\pr}  & = \Gamma^{\omega\nu\nu'}_{r,lmm\pr l\pr} + \phi^{\omega\nu\nu'}_{r,lmm\pr l\pr},\label{eqn:floc}\\
\phi^{\omega\nu\nu'}_{r,lmm\pr l\pr} & = \sum_{\substack{nn\pr hh\pr \\ \nu''}}\Gamma^{\omega\nu\nu''}_{r,lmhn} \chi^{\omega\nu''\nu''}_{0,nhh\pr n\pr} F^{\omega\nu''\nu'}_{r,n\pr h\pr m\pr l\pr}\label{eqn:philoc}.
\end{align}
Here $r\in\{d,m\}$ denotes the (d)ensity or the (m)agnetic spin combination as in Eqs.\ (\ref{eqn:d}) and (\ref{eqn:m}) which allows us to decouple the spin components. The local full vertex $F$ can in turn be obtained via Eq.~(\ref{eqn:gfromf}) from the local 2-particle Green's function $G^{{\mathrm{con~}}}$, which can be directly calculated in continuous-time quantum Monte Carlo. 
Equivalently, $\Gamma$ can also be directly obtained from the local BSE of local generalized susceptibilities, 
\begin{equation}
\chi^{\omega\nu\nu'}_{r,lmm\pr l\pr}  = \chi^{\omega\nu\nu'}_{0,lmm\pr l\pr}\delta_{\nu\nu\pr} 
+ \sum_{\substack{nn\pr hh\pr \\ \nu''}}\chi^{\omega\nu\nu}_{0,lmhn} \Gamma^{\omega\nu\nu''}_{r,nhh\pr n\pr} \chi^{\omega\nu''\nu'}_{r,n\pr h\pr m\pr l\pr},
\label{eq:locchi}
\end{equation}
as depicted in \fref{Fig:flow}(a).

The BSE extends the ``swapping'' symmetry in Eq.~(\ref{eqn:swap}) of $F$ to $\phi$ and $\Gamma$, but not the crossing symmetry in Eqs.~(\ref{eqn:cross1}) and (\ref{eqn:cross2}), i.e.,
\begin{align}
\phi^{\omega\nu\nu'}_{r,lmm\pr l\pr} & = \phi^{(-\omega)(\nu'-\omega)(\nu-\omega)}_{r,m\pr l\pr lm},\\
\phi^{\omega\nu\nu'}_{r,lmm\pr l\pr} & \neq \phi^{\omega\nu\nu'}_{\overline{ph},r,lmm\pr l\pr},\\
\phi^{\omega\nu\nu'}_{\overline{ph},\sigma\sigma',lmm\pr l\pr} & = -\phi^{(\nu'-\nu)(\nu'-\omega)\nu'}_{\overline{\sigma'\sigma},m\pr mll\pr} \\
& = -\phi^{(\nu-\nu')\nu(\nu-\omega)}_{\overline{\sigma'\sigma},ll\pr m\pr m},\label{eqn:phiphbar}
\end{align}
where the transversal particle-hole channel ($\overline{ph}$) by definition is antisymmetric to the particle-hole channel with respect to a relabelling of the two incoming or outgoing particles. Applying the SU(2) symmetry relations in Eq.~(\ref{eqn:su2}) to $\phi_{\overline{ph}}$ gives the explicit relations
\begin{align}
\phi^{\omega\nu\nu'}_{\overline{ph},d,lmm\pr l\pr} & = -\frac{1}{2}\phi^{(\nu'-\nu)(\nu'-\omega)\nu'}_{d,m\pr m ll\pr} - \frac{3}{2}\phi^{(\nu'-\nu)(\nu'-\omega)\nu'}_{m,m\pr m ll\pr},\label{eqn:phidbar}\\
\phi^{\omega\nu\nu'}_{\overline{ph},m,lmm\pr l\pr} & = -\frac{1}{2}\phi^{(\nu'-\nu)(\nu'-\omega)\nu'}_{d,m\pr m ll\pr} + \frac{1}{2}\phi^{(\nu'-\nu)(\nu'-\omega)\nu'}_{m,m\pr m ll\pr},\label{eqn:phimbar}
\end{align}
or in the case of a non-local BSE
\begin{align}
\phi^{\cvec{q}\cvec{k}\cvec{k}'}_{\overline{ph},d,lmm\pr l\pr} & = -\frac{1}{2}\phi^{(\cvec{k}'-\cvec{k})(\cvec{k}'-\cvec{q})\cvec{k}'}_{d,m\pr m ll\pr} - \frac{3}{2}\phi^{(\cvec{k}'-\cvec{k})(\cvec{k}'-\cvec{q})\cvec{k}'}_{m,m\pr m ll\pr},\label{eqn:phidqbar}\\
\phi^{\cvec{q}\cvec{k}\cvec{k}'}_{\overline{ph},m,lmm\pr l\pr} & = -\frac{1}{2}\phi^{(\cvec{k}'-\cvec{k})(\cvec{k}'-\cvec{q})\cvec{k}'}_{d,m\pr m ll\pr} + \frac{1}{2}\phi^{(\cvec{k}'-\cvec{k})(\cvec{k}'-\cvec{q})\cvec{k}'}_{m,m\pr m ll\pr}.\label{eqn:phimqbar}
\end{align}

From the starting point $\Gamma^{\cvec{q}\cvec{k}\cvec{k}'}_{r}$ in Eq.~(\ref{eqn:gammastart}), we now need to construct the full vertex $F^{\cvec{q}\cvec{k}\cvec{k}'}_r$ through a non-local BSE. In the following we will focus on the longitudinal particle-hole channel, but the final expressions will also contain the BSE diagrams for the transversal particle-hole channel through the use of Eqs.(\ref{eqn:phidqbar}) and (\ref{eqn:phimqbar}). The third channel, the particle-particle channel, is considered here to be local in nature and already well described by its local contribution in $\Gamma^{\omega\nu\nu'}$. 

The non-local BSE in the particle-hole channel is given by
\begin{align}
F^{\cvec{q}\cvec{k}\cvec{k}'}_{r,lmm\pr l\pr}  & = \Gamma^{\cvec{q}\cvec{k}\cvec{k}'}_{r,lmm\pr l\pr} + \sum_{\substack{nn\pr hh\pr \\ \cvec{k}''}}\Gamma^{\cvec{q}\cvec{k}\cvec{k}''}_{r,lmhn} \chi^{\cvec{q}\cvec{k}''\cvec{k}''}_{0,nhh\pr n\pr} F^{\cvec{q}\cvec{k}''\cvec{k}'}_{r,n\pr h\pr m\pr l\pr}. \label{eqn:BSEnonlocal}
\end{align}
A considerable simplification of this equation 
is possible if  $\Gamma^{\cvec{q}\cvec{k}\cvec{k}'}$
does not depend on the momenta $\kvec{k}$ and $\kvec{k}'$. Indeed this dependence arises only from the  second (crossed)
$V^{\kvec{k}'-\kvec{k}}$ term in Eq.\ (\ref{eqn:vqkk}) which is neglected e.g.\ in the $GW$ approach. If we follow $GW$ and neglect this term or average it over $\kvec{k}$ (which gives zero since $V$ was defined as purely non-local), the vertex (now already in the two spin channels $r\in\{d,m\}$) reads
\begin{equation}
\Gamma^{\cvec{q}\nu\nu'}_{r,lmm\pr l\pr}  = \Gamma^{\omega\nu\nu'}_{r,lmm\pr l\pr} + 2\beta^{-2} V^{\kvec{q}}_{lm\pr ml\pr} \delta_{r,d}.\label{eqn:gammavd} 
\end{equation}
and the BSE becomes (see  Fig.~\ref{Fig:flow}(c))
\begin{equation}
F^{\cvec{q}\cvec{k}\cvec{k}'}_{r,lmm\pr l\pr}  = \Gamma^{\cvec{q}\nu\nu'}_{r,lmm\pr l\pr} + \sum_{\substack{nn\pr hh\pr \\ \cvec{k}''}}\Gamma^{\cvec{q}\nu\nu''}_{r,lmhn} \chi^{\cvec{q}\cvec{k}''\cvec{k}''}_{0,nhh\pr n\pr} F^{\cvec{q}\cvec{k}''\cvec{k}'}_{r,n\pr h\pr m\pr l\pr}.\label{eqn:prefqnunu}
\end{equation}
Since $\Gamma_r$ is now independent of $\kvec{k}$ and $\kvec{k}'$, this will also be the case for $F$ in Eq.~(\ref{eqn:prefqnunu}). The summation over $\kvec{k}''$ hence  yields
\begin{align}
F^{\cvec{q}\nu\nu'}_{r,lmm\pr l\pr}  & = \Gamma^{\cvec{q}\nu\nu'}_{r,lmm\pr l\pr} + \phi^{\cvec{q}\nu\nu'}_{r,lmm\pr l\pr}\label{eqn:fqnunu}\\
\phi^{\cvec{q}\nu\nu'}_{r,lmm\pr l\pr} & = \sum_{\substack{nn\pr hh\pr \\ \nu''}}\Gamma^{\cvec{q}\nu\nu''}_{r,lmhn} \chi^{\cvec{q}\nu''\nu''}_{0,nhh\pr n\pr} F^{\cvec{q}\nu''\nu'}_{r,n\pr h\pr m\pr l\pr},\label{eqn:phinonloc}\\ 
\chi^{\cvec{q}\nu\nu}_{0,lmm\pr l\pr} & = \sum_{\kvec{k}} \chi^{\cvec{q}\cvec{k}\cvec{k}}_{0,lmm\pr l\pr}. \label{eq:bubble_q}
\end{align}
By combining the left (right) orbital indices and fermionic Matsubara frequencies into a single compound index $\{lm,\nu \}$ ($\{l\pr m\pr,\nu' \}$) Eq.~(\ref{eqn:fqnunu}) can be written as a matrix equation in terms of these compound indices:
\begin{align}
F^{\cvec{q}}_{r} & = \Gamma^{\cvec{q}}_{r} + \phi^{\cvec{q}}_r =\Gamma^{\cvec{q}}_{r} + \Gamma^{\cvec{q}}_{r} \chi^{\cvec{q}}_{0} F^{\cvec{q}}_{r}.\label{eqn:fqmatrix}
\end{align}
The full vertex $F$ can now, in principle, be extracted from Eq.~(\ref{eqn:fqmatrix}) through a simple matrix inversion
\begin{equation}
F^{\cvec{q}}_{r} = \left[ (\Gamma^{\cvec{q}}_{r})^{-1} - \chi^{\cvec{q}}_{0} \right]^{-1}.\label{eqn:f}
\end{equation}
However, as recently shown in Ref.\ \onlinecite{Schaefer2016b} the local $\Gamma$ extracted from a self-consistent DMFT calculation contains an infinite set of diverging components. The numerical complications associated with these diverging components can be avoided by substituting the local $\Gamma$  in Eq.~(\ref{eqn:f}) by the local $F$ using Eqs.~(\ref{eqn:floc}) and (\ref{eqn:philoc}). After some algebra this yields 
\begin{align}
F^{\cvec{q}}_{d} = & \Big( F^{\omega}_d + 2\beta^{-2} V^{\kvec{q}}(\idmatrix + \chi^{\omega}_{0} F^{\omega}_d) \Big) \times \nonumber\\ 
& \Big[\idmatrix - \chi^{nl,\cvec{q}}_{0} F^{\omega}_d - 2\beta^{-2}\chi^{\cvec{q}}_{0}V^{\kvec{q}}(\idmatrix + \chi^{\omega}_{0} F^{\omega}_d)  \Big]^{-1}\!\!\!\!\!\!\!,\label{eqn:fvd}\\
F^{\cvec{q}}_{m} = & F^{\omega}_m \Big[\idmatrix - \chi^{nl,\cvec{q}}_{0} F^{\omega}_m \Big]^{-1} \label{eqn:fvm}
\end{align}
where the purely non-local $\chi^{nl}$ is defined as
\begin{equation}
\chi^{nl,\cvec{q}}_{0} \equiv \chi^{\cvec{q}}_{0} - \chi^{\omega}_{0}.\label{eqn:chinl}
\end{equation}
This formulation is equivalent to Eq.~(\ref{eqn:f}) but circumvents the aforementioned divergencies in the local $\Gamma$.

The non-local full vertices generated in Eqs.~(\ref{eqn:fvd}) and (\ref{eqn:fvm}) through only the particle-hole channel are not crossing symmetric [the vertices are not antisymmetric with respect to a relabelling of the two in-coming or outgoing particles, as in Eqs.~(\ref{eqn:cross1}) and (\ref{eqn:cross2})]. The crossing symmetry is  however restored if we take the corresponding diagrams in the transversal particle-hole channel into account as well,  as done before for a single orbital \cite{Toschi07,Katanin2009}. That is, in the parquet equation we add the reducible contributions in the particle-hole and transversal particle-hole channel
and subtract their respective local contribution which is already contained in the local $F$:
\begin{align}
{\mathbf{F}}^{\cvec{q}\cvec{k}\cvec{k}'}_{d,lmm\pr l\pr} = & F^{\omega\nu\nu'}_{d,lmm\pr l\pr} + {\mathbf{V}}^{\kvec{q}\kvec{k}\kvec{k}'}_{d,lmm\pr l\pr} + (\phi^{\cvec{q}\nu\nu'}_{d,lmm\pr l\pr} - \phi^{\omega\nu\nu'}_{d,lmm\pr l\pr}) \nonumber\\
 & + (\phi^{\cvec{q}\cvec{k}\cvec{k}'}_{\overline{ph},d,lmm\pr l\pr} - \phi^{\omega\nu\nu'}_{\overline{ph},d,lmm\pr l\pr})\label{eqn:parquet} .
\end{align}
Here, we consider the particle-particle channel and all fully irreducible diagrams, except ${\mathbf{V}}^{\kvec{q}\kvec{k}\kvec{k}'}$, to be local. The bare non-local interaction vertex ${\mathbf{V}}^{\kvec{q}\kvec{k}\kvec{k}'}$ defined in Eq.~(\ref{eqn:vqkk}) has to be added explicitly to the parquet equation since it is neither part of the reducible vertices $\phi_{ph}$ and $\phi_{\overline{ph}}$, nor the local $F$.

Resolving Eqs.~(\ref{eqn:floc}) and (\ref{eqn:fqnunu}) for $\phi$, Eq.~(\ref{eqn:gammavd}) for $\Gamma$, and taking the difference of the local and non-local $\phi$ yields 
\begin{align}
(\phi^{\cvec{q}\nu\nu'}_{d,lmm\pr l\pr} - \phi^{\omega\nu\nu'}_{d,lmm\pr l\pr}) = & ~F^{nl,\cvec{q}\nu\nu'}_{d,lmm\pr l\pr} - 2 \beta^{-2}V^{\kvec{q}}_{lm\pr ml\pr},\label{eqn:phiphid}
\end{align}
where the (full) non-local vertex $F^{nl}$ is defined as
\begin{equation}
F^{nl,\cvec{q}\nu\nu'}_{r,lmm\pr l\pr} \equiv F^{\cvec{q}\nu\nu'}_{r,lmm\pr l\pr} - F^{\omega\nu\nu'}_{r,lmm\pr l\pr}.\label{eqn:fnl}
\end{equation}
For the transversal particle-hole channel we can calculate the same difference 
by subtracting   Eq.\ (\ref{eqn:phidbar}) from   Eq.\ (\ref{eqn:phidqbar}) and expressing all terms by $F$ similar as in Eq.\ (\ref{eqn:phiphid}). This yields 
\begin{align}
(\phi^{\cvec{q}\cvec{k}\cvec{k}'}_{\overline{ph},d,lmm\pr l\pr} - \phi^{\omega\nu\nu'}_{\overline{ph},d,lmm\pr l\pr}) = & - \frac{1}{2} F^{nl,(\cvec{k}'-\cvec{k})(\nu'-\omega)\nu'}_{d,m\pr m ll\pr}\nonumber\\
& -  \frac{3}{2} F^{nl,(\cvec{k}'-\cvec{k})(\nu'-\omega)\nu'}_{m,m\pr mll\pr} \nonumber\\
& +  \beta^{-2} V^{\kvec{k}'-\kvec{k}}_{m\pr lml\pr}.\label{eqn:phiphidbar}
\end{align}

 Eqs.~(\ref{eqn:phiphid}) and (\ref{eqn:phiphidbar}) can now be used in Eq.\ (\ref{eqn:parquet}) to finally give
\begin{align}
{\mathbf{F}}^{\cvec{q}\cvec{k}\cvec{k}'}_{d,lmm\pr l\pr} = & F^{\omega\nu\nu'}_{d,lmm\pr l\pr} + F^{nl,\cvec{q}\nu\nu'}_{d,lmm\pr l\pr} - \frac{1}{2} F^{nl,(\cvec{k}'-\cvec{k})(\nu'-\omega)\nu'}_{d,m\pr mll\pr} \nonumber\\
&- \frac{3}{2}F^{nl,(\cvec{k}'-\cvec{k})(\nu'-\omega)\nu'}_{m,m\pr mll\pr},\label{eqn:ffinal}
\end{align}
where the non-local $F^{nl}$ is defined in Eq.\ (\ref{eqn:fnl}) with  $F^{\cvec{q}}$ from the reformulated BSEs (\ref{eqn:fvd}) or (\ref{eqn:fvm}).

It should be noted that the two non-crossing symmetric contributions to the bare non-local interaction $V$ in Eq.~(\ref{eqn:phiphid}) and (\ref{eqn:phiphidbar}) add up to become exactly $V^{\kvec{q}\kvec{k}\kvec{k}'}$ as defined in Eq.~(\ref{eqn:vqkk}). This is unique to the simplification employed in Eq.~(\ref{eqn:gammavd}). 

\subsection{Equation of motion}
\label{EoM}

Besides the BSE, the equation of motion or Schwinger-Dyson equation is the second central equation of the AbinitioD$\Gamma$A approach. It allows us to calculate the self-energy from the crossing symmetric full vertex (or the connected two-particle Green's function).  For deriving the multi-orbital Schwinger-Dyson equation, we
compare the $\tau$-derivative of $ G^{\kvec{k}}_{\sigma,lm}(\tau)$ in the Heisenberg equation of motion with the Dyson equation. This yields 
\begin{align}
\MoveEqLeft {\left[ \Sigma G \right]^{\kvec{k}}_{\sigma,mm\pr}(\tau)  =  \torder{ \left[\op{U}^{{\mathrm{full}}},\op{c}^{\phantom\dagger}_{\kvec{k} m\sigma }(\tau)\right] \op{c}^{\dagger}_{\svek{k} m\pr\sigma }(0) }} & \hfill\nonumber\\
&= \sum_{\substack{ lhn \sigma' \\ \kvec{q}\kvec{k}' }} \left( U_{mlhn} + \mathbf{V}^{\kvec{q}}_{mlhn}\right) \times\nonumber \\ & \times \torder{ \op{c}^{\dagger}_{ \kvec{k}'-\kvec{q}  l\sigma' }(\tau) \op{c}^{\phantom\dagger}_{\kvec{k}-\kvec{q} h\sigma }(\tau)  \op{c}^{\phantom\dagger}_{\kvec{k}' n\sigma' }(\tau) \op{c}^{\dagger}_{\kvec{k}   m\pr\sigma }(0) } \nonumber\\
&= \!\lim_{\tau' \rightarrow \tau^{+}} \! \sum_{\substack{ lhn \sigma' \\ \kvec{q}\kvec{k}' }} \left( U_{mlhn} + \mathbf{V}^{\kvec{q}}_{mlhn}\right) G^{\kvec{q}\kvec{k}'\kvec{k}}_{\sigma'\sigma, nlhm\pr}(\tau,\tau',\tau)\label{eqn:preeom}
\end{align}
where, in the second line, we have used the swapping symmetry for $U_{lm\pr ml\pr}$ and $\mathbf{V}^{\kvec{q}}_{lm\pr ml\pr}$. The limit in Eq.~(\ref{eqn:preeom}) can be taken by splitting the two-particle Green's function into its connected and disconnected parts using Eq.~(\ref{eqn:gcon}):
\begin{align}
\MoveEqLeft{\left[ \Sigma G \right]^{\kvec{k}}_{\sigma,mm\pr}(\tau) = \sum_{\substack{ lhn \sigma' \\ \kvec{q}\kvec{k}' }} \left( U_{mlhn} + \mathbf{V}^{\kvec{q}}_{mlhn}\right) \times} \nonumber \\
&\times \left[ G^{{\mathrm{con~}} \kvec{q}\kvec{k}'\kvec{k}}_{\sigma'\sigma, nlhm\pr}(\tau,\tau,\tau)   +\delta_{\kvec{q}0}n^{\kvec{k}'}_{\sigma',ln}G^{\kvec{k}}_{\sigma',hm\pr}(\tau)   \right. & \nonumber\\
&\;\;\;\;\left.   - \delta_{\sigma\sigma'}\delta_{\kvec{k}\kvec{k}'}n^{\kvec{k}-\kvec{q}}_{\sigma,lh}G^{\kvec{k}}_{\sigma,nm\pr}(\tau) \right],
\end{align}
where $n_{mm\pr} = \langle \op{c}^{\dagger}_{m } \op{c}^{\phantom\dagger}_{m\pr}\rangle$.
Taking the Fourier transform with respect to $\tau$ gives
\begin{align}
\MoveEqLeft\left[ \Sigma G \right]^{\cvec{k}}_{\sigma,mm\pr} = \sum_{\substack{ lhh\pr n \sigma' \\ \kvec{q}\kvec{k}' }} \left( U_{mlhn} + \mathbf{V}^{\kvec{q}}_{mlhn}\right) \nonumber\\ & \times \left[\int_0^\beta  e^{i\nu\tau} G^{{\mathrm{con~}} \kvec{q}\kvec{k}'\kvec{k}}_{\sigma'\sigma, nlh\pr m\pr}(\tau,\tau,\tau) d\tau \right. \nonumber\\
& \;\;\;\;\left. +\delta_{\kvec{q}0}n^{\kvec{k}'}_{\sigma',ln}G^{\cvec{k}}_{\sigma',h\pr m\pr}  - \delta_{\sigma\sigma'}\delta_{\kvec{k}\kvec{k}'}n^{\kvec{k}-\kvec{q}}_{\sigma,lh}G^{\cvec{k}}_{\sigma,nm\pr} \right].\label{eqn:eomtau}
\end{align}
Since the connected part is continuous it is possible to obtain the equal time component in Eq.~(\ref{eqn:eomtau}) by simply summing up the bosonic and the left fermionic Matsubara frequencies
\begin{equation}
\int_0^\beta d\tau  e^{i\nu\tau} G^{{\mathrm{con~}} \kvec{q}\kvec{k}'\kvec{k}}_{\sigma'\sigma, nlh\pr m\pr}(\tau,\tau,\tau) = \frac{1}{\beta^2}\sum_{\omega\nu'} G^{{\mathrm{con~}} \cvec{q}\cvec{k}'\cvec{k}}_{\sigma'\sigma,nlh\pr m\pr}.\label{eqn:gconsum}
\end{equation}
Finally, multiplying with $G^{-1}$ from the right yields the multi-orbital Schwinger-Dyson equation
\begin{align}
\Sigma^{\cvec{k}}_{\sigma,mm\pr} & = \Sigma^{{\mathrm{HF~}}\kvec{k}}_{\sigma,mm\pr} + \Sigma^{{\mathrm{con~}}\cvec{k}}_{\sigma,mm\pr},\\
\Sigma^{{\mathrm{con~}}\cvec{k}}_{\sigma,mm\pr} & = \beta^{-2}\sum_{\substack{ ll\pr hn \\ \sigma'\cvec{q}\cvec{k}'}} \! \Big( U_{mlhn} + \mathbf{V}^{\kvec{q}}_{mlhn}\Big) G^{{\mathrm{con~}} \cvec{q}\cvec{k}'\cvec{k}}_{\sigma'\sigma, nlhl\pr} [G^{\cvec{k}}_{\sigma}]^{-1}_{l\pr m\pr},\label{eqn:sigmacon}
\end{align}
where $\Sigma^{{\mathrm{HF}}}$ is the static Hartree-Fock contribution to the self-energy. 

Since we would like to calculate the self-energy starting from ${{F}}$ in Eq.~(\ref{eqn:ffinal}), let us recall that we assume SU(2) symmetry and apply the relation between ${{F}}$ and $G^{{\mathrm{con}}}$   in Eq.~(\ref{eqn:gfromf}). This yields the multi-orbital Schwinger-Dyson equation
\begin{align}\label{eqn:sigmafromf}
\Sigma^{{\mathrm{con~}}\cvec{k}}_{mm\pr} = -\beta^{-1} \sum_{\substack{ ll\pr nn\pr hh\pr \\ \cvec{q}\cvec{k}'}} \! &\Big( U_{mlhn} + \mathbf{V}^{\kvec{q}}_{mlhn}\Big)\times \\ \nonumber & \times \chi^{\cvec{q}\cvec{k}'\cvec{k}'}_{0, nll\pr n\pr} F^{\cvec{q}\cvec{k}'\cvec{k}}_{d,n\pr l\pr h\pr m\pr} G^{\cvec{k}-\cvec{q}}_{hh\pr},
\end{align}
that finally determines the non-local AbinitioD$\Gamma$A self-energy.

In the following we present some implementational details.
That is, we split Eq.~(\ref{eqn:sigmafromf}) into contributions of the 
 particle-hole and transversal particle-hole terms of Eq.\ (\ref{eqn:ffinal}) as well into the  $U$ and  $\mathbf{V}^{\kvec{q}}$ terms. This yields, suppressing the orbital indices for clarity as they remain identical to those in Eq.~(\ref{eqn:sigmafromf}):
\begin{align}
\Sigma^{Uloc,\cvec{k}} = & -\beta^{-1} \sum_{\cvec{q}\nu'} \! U \chi^{\cvec{q}\nu'\nu'}_{0} F^{\omega\nu'\nu}_{d} G^{\cvec{k}-\cvec{q}},\\
\Sigma^{Vloc,\cvec{k}} = & -\beta^{-1} \sum_{\cvec{q}\nu'} \! \mathbf{V}^{\kvec{q}} \chi^{\cvec{q}\nu'\nu'}_{0} F^{\omega\nu'\nu}_{d} G^{\cvec{k}-\cvec{q}},\\
\Sigma^{ph,\cvec{k}} = & -\beta^{-1} \sum_{\cvec{q}\nu'} \! \Big( U + \mathbf{V}^{\kvec{q}}\Big) \chi^{\cvec{q}\nu'\nu'}_{0} F^{nl,\cvec{q}\nu'\nu}_{d} G^{\cvec{k}-\cvec{q}},\\
\Sigma^{U\overline{ph},\cvec{k}}  = &  \beta^{-1} \sum_{\cvec{q}\nu'} \! \tilde{U} \chi^{\cvec{q}\nu'\nu'}_{0}\times \\ \nonumber & \times \Big(\frac{1}{2} F^{nl,\cvec{q}\nu'\nu}_{d} + \frac{3}{2} F^{nl,\cvec{q}\nu'\nu}_{m} \Big)G^{\cvec{k}-\cvec{q}},\\
\Sigma^{V\overline{ph},\cvec{k}}  = &  \beta^{-1} \sum_{\cvec{q}\nu'} \! \tilde{V}^{\kvec{k}'-\kvec{k}} \chi^{\cvec{q}\cvec{k}'\cvec{k}'}_{0}\times \\ \nonumber & \times \Big(\frac{1}{2} F^{nl,\cvec{q}\nu'\nu}_{d} + \frac{3}{2} F^{nl,\cvec{q}\nu'\nu}_{m} \Big)G^{\cvec{k}-\cvec{q}},
\end{align}
where $\tilde{U}_{lm\pr l\pr m} = U_{lm\pr ml\pr}$ and similarly for $V$. The indices in the terms originating from the transversal particle-hole channel have been relabelled to make the full vertices $F$ depend on $\kvec{q}$ instead of $\kvec{k}'-\kvec{k}$. 
Indeed, the way Eq.~(\ref{eqn:sigmafromf}) is written might suggest that the particle-hole and transversal particle-hole channels are treated differently. This is however not the case since an application of the crossing symmetry of $F$ together with the swapping symmetry of the interaction leaves Eq.~(\ref{eqn:sigmafromf}) unchanged, but swaps the role of the particle-hole and transversal particle-hole channels in $F$.  
In the BSE ladders we have, in Eq.\ (\ref{eqn:gammavd}) and similar to $GW$, included  $\mathbf{V}^\kvec{q}$ but not  $\mathbf{V}^{\kvec{k}'-\kvec{k}}$. Against this background,  it is reasonable to omit $\Sigma^{V\overline{ph}}$ for consistency.

 In the following we will take advantage of the particular momentum and frequency structure of the Schwinger-Dyson equation to optimize the numerical calculation of the self-energy. To this end we define  three three-legged quantities (cf.\ Refs. \onlinecite{Katanin2009,Ayral2015}) with  increasing order of non-local character:
\begin{align}
\gamma^{\omega\nu}_{r,lmm\pr l\pr} & \equiv \sum_{n\pr h\pr \nu'} \chi^{\omega\nu'\nu'}_{0,lmn\pr h\pr} F^{\omega\nu'\nu}_{r,h\pr n\pr m\pr l\pr},\\
\gamma^{\cvec{q}\nu}_{r,lmm\pr l\pr} & \equiv \sum_{n\pr h\pr\nu'} \chi^{nl,\cvec{q}\nu'\nu'}_{0,lmn\pr h\pr} F^{\omega\nu'\nu}_{r,h\pr n\pr m\pr l\pr},\\
\eta^{\cvec{q}\nu}_{r,lmm\pr l\pr} & \equiv \sum_{n\pr h\pr\nu'} \chi^{\cvec{q}\nu'\nu'}_{0,lmn\pr h\pr} F^{\cvec{q}\nu'\nu}_{r,h\pr n\pr m\pr l\pr} - \chi^{\omega\nu'\nu'}_{0,lmn\pr h\pr} F^{\omega\nu'\nu}_{r,h\pr n\pr m\pr l\pr}.
\end{align}
Here, $\gamma^{\omega\nu}$ is strictly local and can be extracted directly from the impurity solver\cite{Hafermann12,Gunacker16}; $\gamma^{\cvec{q}\nu}$ contains the local full vertex connected to a purely non-local bare two-particle propagator. The vertex $\eta^{\cvec{q}\nu}$ describes the full vertex connected to the bare two-particle propagator, but with all purely local diagrams removed. It can be calculated efficiently from Eqs.~(\ref{eqn:fvd}) and (\ref{eqn:fvm}) using a matrix inversion and $\gamma^{\omega}_{r}$:
\begin{align}
\eta^{\cvec{q}}_{r} &= (\vec{\idmatrix} + \gamma^{\omega}_{r})\times\label{eqn:eta} \\ &\times\Big( \Big[ \idmatrix - \chi^{nl,\cvec{q}}_0 F^{\omega}_r - 2\beta^{-2}\chi^{\cvec{q}}_{0}\mathbf{V}^{\kvec{q}}(\vec{\idmatrix} + \gamma^{\omega}_{r})\delta_{rd} \Big]^{-1}-\idmatrix\Big) \nonumber ,
\end{align}
where $\vec{\idmatrix}_{lmm\pr l\pr} = \delta_{ll\pr}\delta_{mm\pr}$. The self-energy can now be written in terms of $\gamma$ and $\eta$.
\begin{align}
\Sigma^{Uloc,\cvec{k}} = & \Sigma^{\nu}_{{\mathrm{DMFT}}}  -\beta^{-1}\sum_{\cvec{q}} \! U \gamma^{\cvec{q}}_{d} G^{\cvec{k}-\cvec{q}},\\
\Sigma^{Vloc,\cvec{k}} = & -\beta^{-1} \sum_{\cvec{q}} \! \mathbf{V}^{\kvec{q}} (\gamma^{\cvec{q}}_{d} + \gamma^{\omega}_{d}) G^{\cvec{k}-\cvec{q}},\\
\Sigma^{ph,\cvec{k}}  = & -\beta^{-1} \sum_{\cvec{q}} \! \Big( U + \mathbf{V}^{\kvec{q}}\Big) (\eta^{\cvec{q}}_{d} - \gamma^{\cvec{q}}_d )G^{\cvec{k}-\cvec{q}},\\
\Sigma^{U\overline{ph},\cvec{k}}  = &  \beta^{-1} \sum_{\cvec{q}} \! \tilde{U} \Big[\frac{1}{2} (\eta^{\cvec{q}}_{d} - \gamma^{\cvec{q}}_d ) + \frac{3}{2} (\eta^{\cvec{q}}_{m} - \gamma^{\cvec{q}}_m )  \Big]G^{\cvec{k}-\cvec{q}}.
\end{align}
By gathering the terms and using the crossing symmetry of the local $F$ in $\gamma^{\cvec{q}}$, one finally obtains for the AbinitioD$\Gamma$A self-energy
\begin{align}\label{eq:eom_final}
\Sigma^{{\mathrm{D}}\Gamma{\mathrm{A}~}} = & \Sigma^{Uloc,\cvec{k}} + \Sigma^{Vloc,\cvec{k}} + \Sigma^{ph,\cvec{k}} + \Sigma^{U\overline{ph}}
\nonumber\\ = & \Sigma^{\nu}_{{\mathrm{DMFT}}} -\beta^{-1} \sum_{\cvec{q}} \! \Big( U + \mathbf{V}^{\kvec{q}} - \frac{\tilde{U}}{2}\Big)\eta^{\cvec{q}}_{d}G^{\cvec{k}-\cvec{q}} +\!\!\!\!
\nonumber\\ & +\beta^{-1} \sum_{\cvec{q}} \! \frac{3}{2}\tilde{U}\eta^{\cvec{q}}_{m}G^{\cvec{k}-\cvec{q}} -\!\!\!\! \nonumber\\ & -\beta^{-1} \sum_{\cvec{q}} \! \Big( \mathbf{V}^{\kvec{q}}\gamma^{\omega}_{d}- U\gamma^{\cvec{q}}_{d}\Big)G^{\cvec{k}-\cvec{q}}  .
\end{align}
In Sec. \ref{Sec:results}, we will apply this AbinitioD$\Gamma$A algorithm to the testbed material SrVO$_3$.

\subsection{Numerical effort}
Before turning to the results for SrVO$_3$, let us briefly discuss the numerical effort of the method.
The numerical effort for calculating  the local vertex in CT-HYB scales roughly as $\beta^5(\#o)^4$ with a large prefactor because of the Monte-Carlo sampling ($\#o$ is the number of oritals; there is also an exponential scaling in $\#o$ for calculating the local trace but only with  a  $\beta^1$ prefactor so that this term is less relevant for typical $\#o$ and $\beta$). The $\beta^5(\#o)^4$ scaling can be understood from the fact that an update of the hybridization matrix is $\sim  \beta^2$ (the mean expansion order is $\sim \beta$), and we need
to determine $(\beta)^3(\#o)^4$ different vertex contributions if the number of measurements per imaginary time interval stays constant.  However, since we eventually calculate the self energy which depends on only one frequency and two orbitals, a much higher noise level can be permitted for larger $\#\omega$  and  $\#o$. That is,  in practice a weaker scaling on  $(\#\omega)$ and $(\#o)$ is possible. Outside a window of lowest frequencies, one can also employ the asymptotic form\cite{Li2016,Wentzell2016} of the vertex which depends on only  two frequencies so that its calculation scales as  $\beta^4(\#o)^4$.
Without using these shortcuts, calculating the vertex for SrVO$_3$ with $\#o=3$ and $\beta=10\,$eV$^{-1}$ took 150000 core h (Intel Xeon E5-2650v2, 2.6 GHz, 16 cores per node).   

As for the AbinitioD$\Gamma$A calculation of the  non-local Feynman diagrams, 
 a parallelization over the compound index $\cvec{q} = (\omega,\kvec{q})$ is 
suitable since $\cvec{q}$  is an external index in the non-local
 Bethe-Salpeter equation (\ref{eqn:fqmatrix}) and the equation of
 motion (\ref{eq:eom_final}). Obviously, this q-loop scales with 
the number of $\kvec{q}$-points $\#q$ and the number of (bosonic)
 Matsubara frequencies $\#\omega$ (which is roughly $\sim  \beta$), 
and thus as $\#\omega\#q$. Within this parallel loop, the numerically 
most demanding task is the matrix inversion in Eq.\ (\ref{eqn:eta}). Since the dimension of the matrix that needs to
 be inverted is given by 
$\#\omega (\#o)^2$ the inversion scales  $\sim (\#\omega\#o^2)^3$.
 Altogether this part hence scales as  $\#q\#\omega^4\#o^6$. 
(The numerical effort for calculating the self energy via the equation 
of motion  (\ref{eq:eom_final}) is  $\sim\#q^2\#\omega^2\#o^6$ and becomes the leading 
contribution at high temperatures and a large number of $\kvec{q}$-points.) 
For the present AbinitioD$\Gamma$A computation of SrVO$_3$ with $\#o=3$,  $\beta=10\,$eV$^{-1}$ ($\#\omega=120$) and $\#q=20^3$, the total computational effort  of this part was 3200 core h.

\section{Results for {SrVO$_3$}}
\label{Sec:results}

Strontium vanadate, SrVO$_3$, is a strongly correlated metal that crystallizes 
in a cubic perovskite lattice structure with lattice constant $a=\,3.8$\AA.
It has a mass enhancement of $m^*/m\sim 2$ according to photoemission spectroscopy\cite{Sekiyama2004} 
and specific heat measurements\cite{PhysRevB.58.4372}. At low frequencies, SrVO$_3$ further reveals a correlation induced kink in the energy-momentum dispersion relation \cite{Nekrasov05a,Byczuk2007,Aizaki12,Held13} if subject to  careful examination \cite{Aizaki12}.
SrVO$_3$ became the testbed material for the benchmarking of new codes and the testing of new methods for strongly correlated electron systems, see e.g.\ Refs.~\onlinecite{Sekiyama2004,Pavarini04,Nekrasov05a,Lee12,PhysRevB.85.035115,Tomczak2012,Taranto2013,Miyake13,PhysRevB.88.235110,Tomczak14,Kazuma16,Boehnke2016,Roekeghem2014_2}. 
Besides academic interests, SrVO$_3$ actually has a number of potential technological applications, e.g.\ as
electrode material\cite{ADMA:ADMA201300900},  Mott transistor\cite{PhysRevLett.114.246401}, or as a transparent conductor.\cite{Zhang2016}

\begin{figure}[t!]
\centering
\begin{subfigure}{0.24\textwidth}
  \centering
  \includegraphics[width=0.9\textwidth]{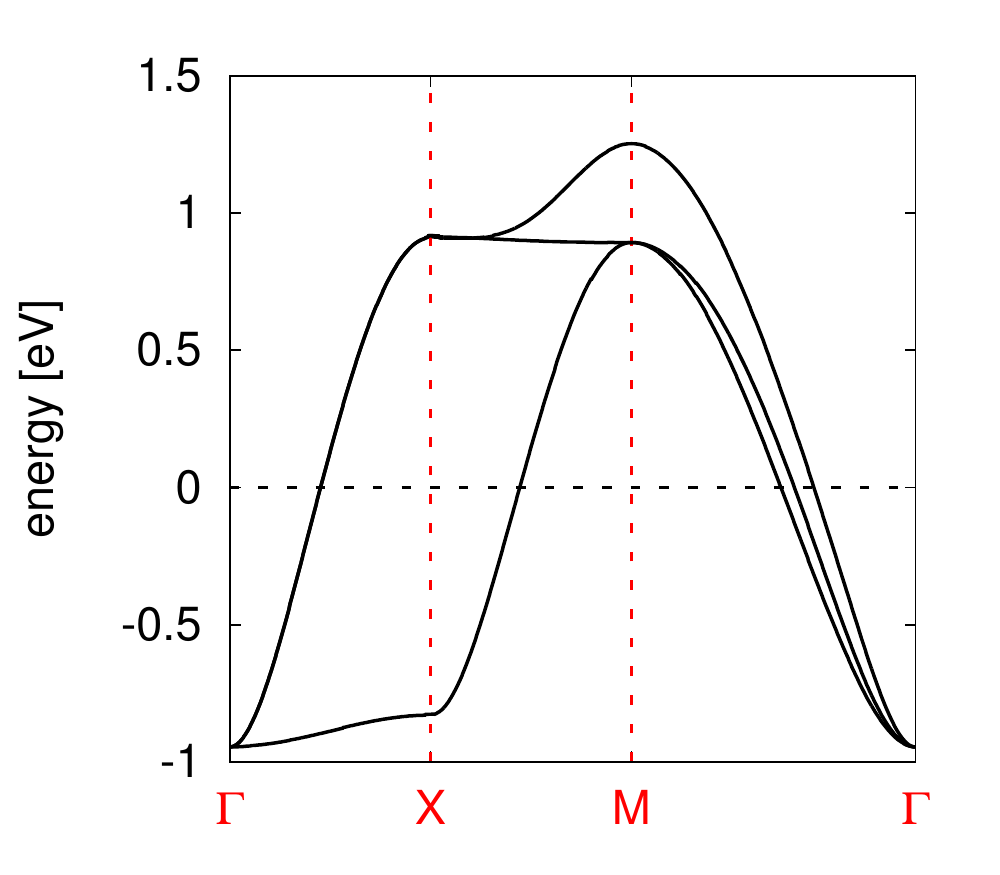}\\[0.55cm]
  \label{fig:fermi_surface}
\end{subfigure}%
\begin{subfigure}{0.24\textwidth}
  \centering
  \includegraphics[width=0.9\textwidth]{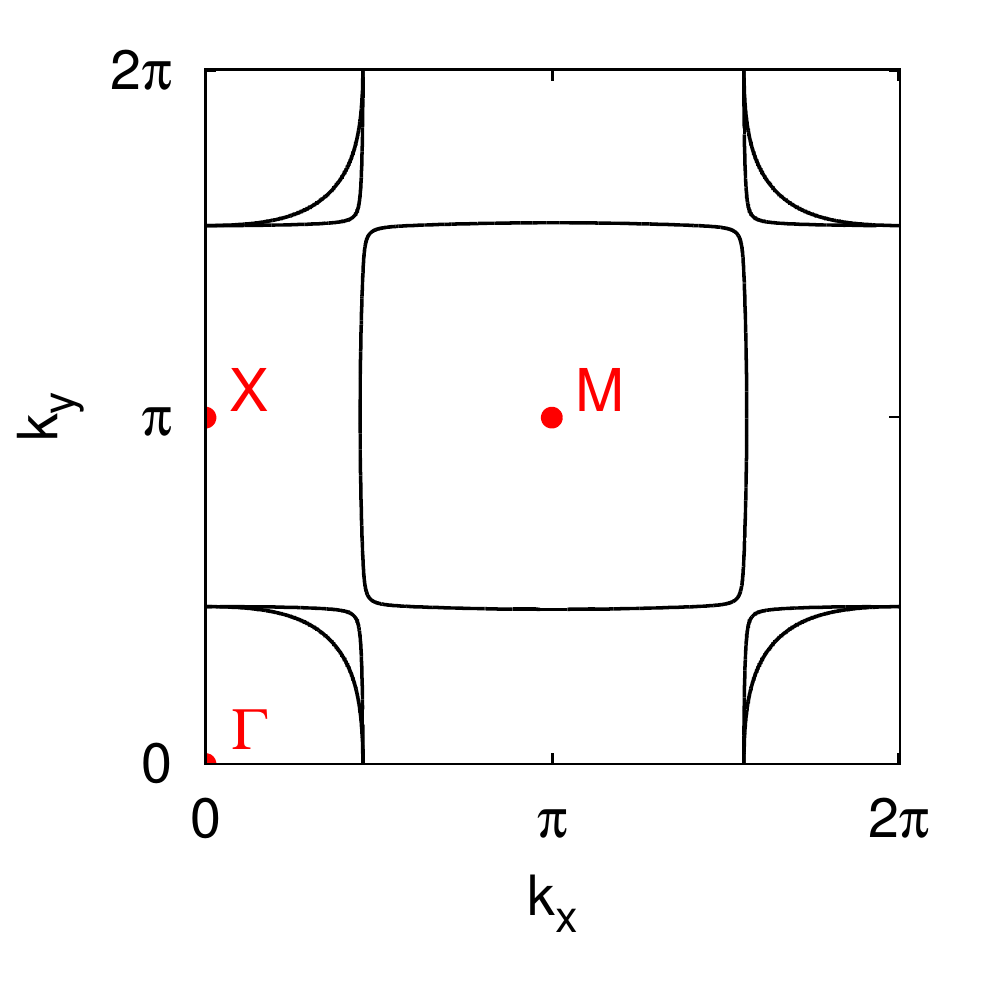}
  \label{fig:bands}
\end{subfigure}
\caption{Bandstructure and Fermi surface of SrVO$_3$ within GGA:
Shown is the dispersion of the vanadium $t_{2g}$ states   (left) and the Fermi surface in the $(k_x,k_y)$-plane for $k_z=0$ (right).}
\label{fig2}
\end{figure}

\begin{figure}
\includegraphics[clip=true,width=9cm,trim={0 1.25cm 0 2cm}]{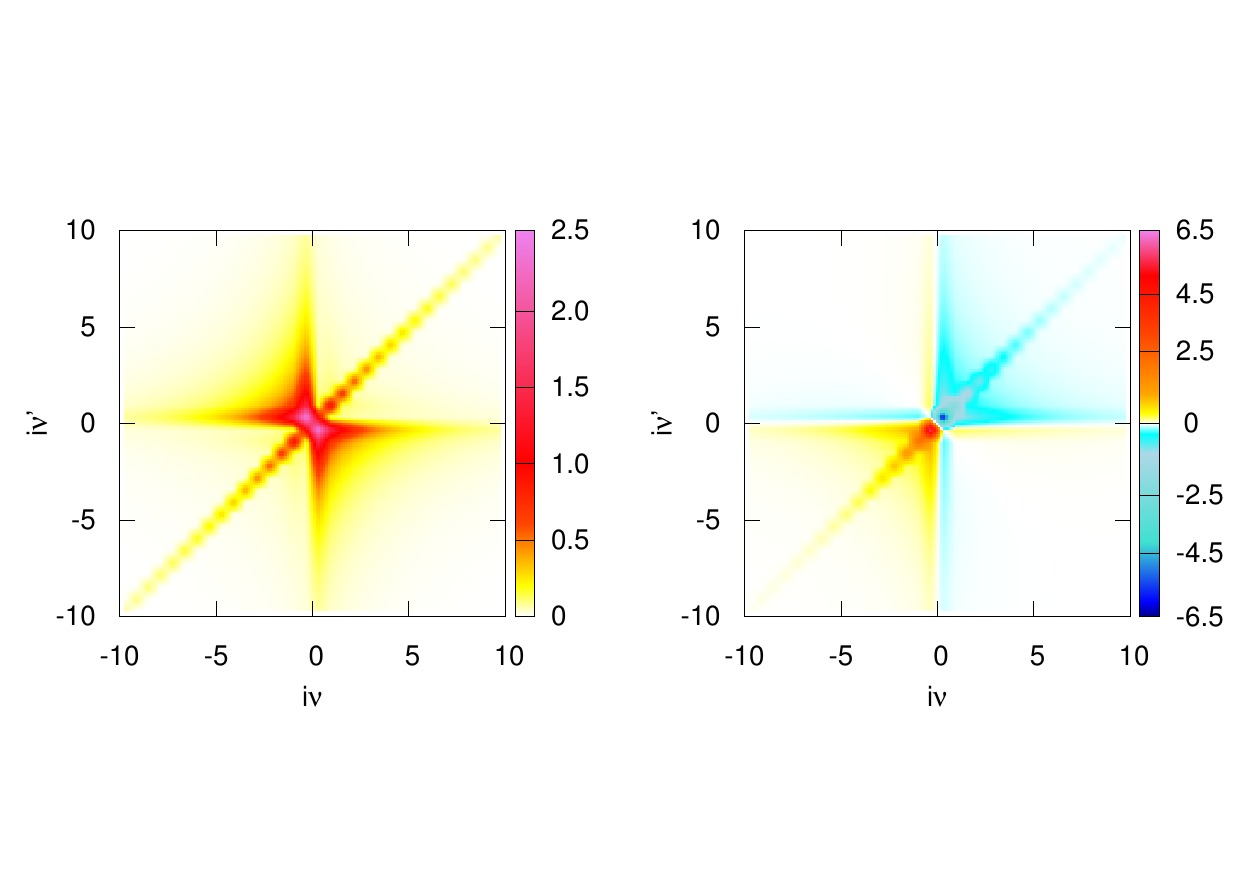}
\caption{(Color online) 
Real (left) and imaginary (right) part of the generalized susceptibility $\chi_{m,1111}^{\omega\nu\nu^\prime}$ in the magnetic ($m$) channel for the $1111$ orbital component at $\omega=0$. 
$\chi$ is related to the irreducible local vertex via Eq.~(\ref{eq:locchi}). By summing $\chi_{m}^{\omega\nu\nu^\prime}$ over its two fermionic frequencies $\nu$ and $\nu^\prime$ one can obtain the physical local magnetic susceptibility $\chi_{m}^{\omega}$, as e.g.\ in Ref.\ \onlinecite{PhysRevB.92.205132}.
\label{fig:vertex_plot}}
\end{figure}

\begin{figure*}[t!]
\begin{minipage}{18 cm}
\includegraphics[clip=true,width=18cm]{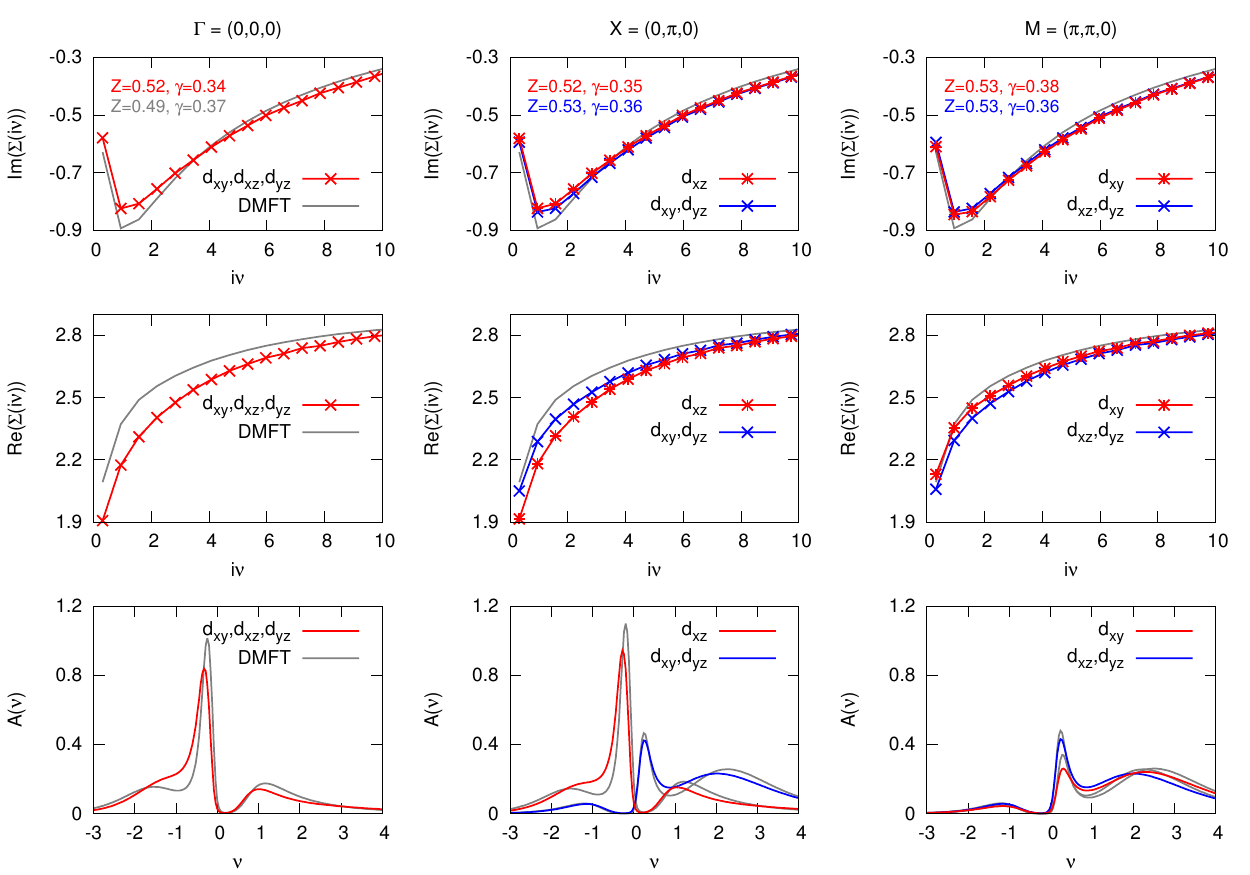}
\caption{(Color online)
 AbinitioD$\Gamma$A $\svek{k}$-dependent self energies and spectral functions for SrVO$_3$. Shown are the imaginary (top) and real  (middle) part of the self-energy and the corresponding spectral function (bottom) for the k-points $\Gamma=(0,0,0)$ (first column), $X=(0,\pi,0)$ (second column) and $M=(\pi,\pi,0)$ (third column).}
\label{fig:siw_multiplot}
\end{minipage}
\end{figure*}

Here we first employ Wien2K \cite{Schwarz2003} bandstructure calculations in the generalized gradient approximation (GGA) \cite{Perdew96} and wien2wannier \cite{Kuneifmmodecheckselsevsfi2010a} to project onto maximally localized Wannier functions\cite{Mostofi2008} for the low energy $t_{2g}$ orbitals of vanadium. 
The momentum dispersion corresponding to these orbitals is shown in Fig.~\ref{fig2} (left) along with a cut of the Fermi surface (right).
For these low-energy orbitals the constrained local density approximation yields an intra-orbital Hubbard $U=5\,$eV, a Hund's exchange $J=0.75\,$eV  and an inter-orbital $U^\prime=U-2J=3.5\,$eV. \cite{Sekiyama2004,Nekrasov05a}
These interaction values were shown to reproduce the experimental mass enhancement within DMFT.\cite{Sekiyama2004,Pavarini04,Nekrasov05a} 

We use the Kanamori parametrisation of the local interaction with the above values for $U$, $U^\prime$ and $J$ and perform
DMFT calculations for the thus defined low-energy model at an inverse temperature $\beta=10 eV^{-1}$.
In DMFT the lattice model is self-consistently mapped onto an auxiliary single Anderson impurity model (SIAM).\cite{Georges1996}
In order to extract the local dynamic four-point vertex function we use the w2dynamics package,\cite{Parragh12,Wallerberger16} which solves the SIAM
using continuous-time quantum Monte Carlo in the hybridisation expansion (CT-HYB).\cite{Werner06,Gull2011a}
When considering non-density-density interactions (such as the Kanamori interaction), the multi-orbital vertex function is only accessible by extending CT-HYB with a worm algorithm\cite{Gunacker15}.
To illustrate the complexity of this quantity, we display in Fig.~\ref{fig:vertex_plot} the generalized susceptibility $\chi_{m,1111}^{\omega\nu\nu^\prime}$ (related to the vertex via Eq.~\ref{eq:locchi}) 
as a function of the two fermionic frequencies  at zero bosonic frequency and all orbital indices being the same.
We sample a cubic frequency box with 120 points in each direction. For relatively high temperatures of $\beta=10 eV^{-1}$ this box is sufficiently large, although we
suggest an extrapolation to an infinite frequency box for the self-energy in Eq.~\ref{eqn:sigmafromf} or the use of high frequency asymptotics\cite{Kunes2011,Li2016,Wentzell2016} for future calculations.
While the CT-HYB algorithm is in principle numerically exact, the four-point vertex function usually suffers from poor statistics due to finite computation times.
In an effort to limit the statistical uncertainties to an acceptable level, we further make use of a sampling method termed ``improved estimators''.\cite{Hafermann12,Gunacker16}
This method redefines Green's function estimators of CT-HYB by employing local versions of the equation of motion, resulting in an improved high-frequency behavior for sampled quantities.

Following the AbinitioD$\Gamma$A approach developed in Sec.~\ref{Sec:method}, we compute the momentum-dependent self-energy $\Sigma_{mm^\prime}(\svek{k},i\nu)$
for SrVO$_3$ in the $t_{2g}$ subspace ($m=xy,xz,yz$).
Here, we employ a one shot  AbinitioD$\Gamma$A with the local vertex  from a DFT+DMFT calculation (using the constrained DFT interaction) as a starting point. Concomitant to the restriction
to the  $t_{2g}$ subspace and the DFT starting point, we do not include  the
 inter-site interaction $\mathbf{V}^{\kvec{q}}$.

Let us note that  recent $GW$+DMFT studies \cite{Tomczak2012,Tomczak14,Boehnke2016} suggest $t_{2g}$ spectral weight above $\sim 1.5$eV to be of plasmonic origin
instead of stemming from the upper Hubbard bands seen in previous (static) DFT+DMFT calculations.
To include this kind of physics one would need to use a frequency dependent $U(\omega)$ from constrained RPA (or a larger window of orbitals in AbinitioD$\Gamma$A taking at least $U$ and $\mathbf{V}^{\kvec{q}}$ as a vertex as discussed in Section
\ref{Sec:method}), as well as non-local interactions $\mathbf{V}^{\kvec{q}}$
 that compete with the bandwidth-narrowing effects from $U(\omega)$ in $GW$+DMFT.\cite{Tomczak14,NoteVSVO}
This goes beyond the scope of the present work, where both aspects are not included, and hence we cannot contribute to this controversy.
Instead, we focus on the 
non-local effects stemming from a local frequency-independent $U$.
These are other corrections to the DFT+DMFT description of  SrVO$_3$.

\begin{figure*}[t!]
\begin{minipage}{18 cm}
\includegraphics[clip=true,width=18cm]{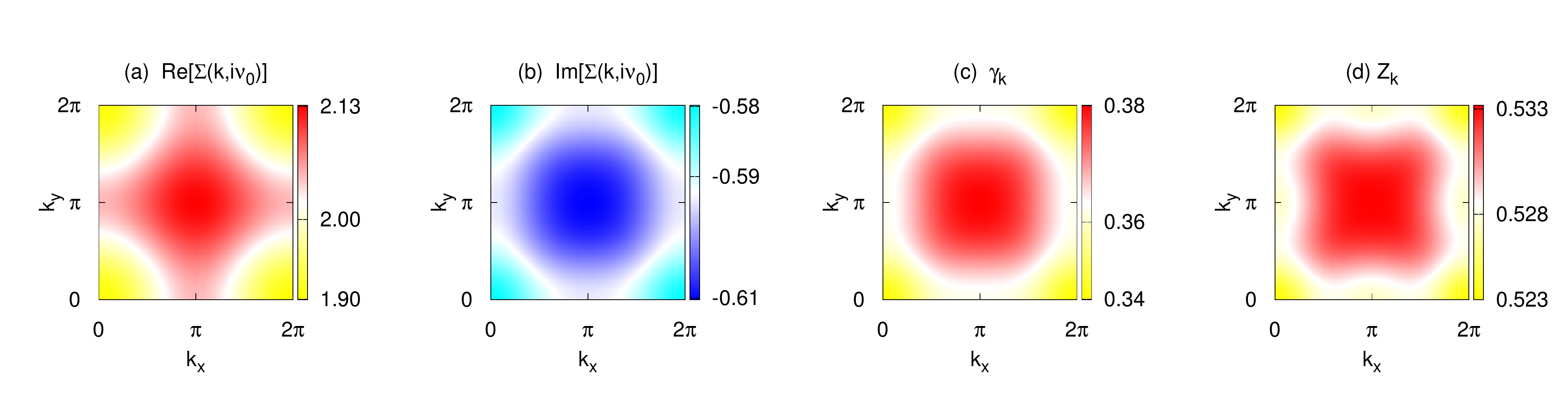}
\caption{(Color online) 
(a) Real  and (b) imaginary part of the AbinitioD$\Gamma$A self-energy $\Sigma(\svek{k},i\nu_0)$ at the first Matsubara frequency $\nu_0$ 
(c)  scattering rate $\gamma_{\svek{k}}$ and (d) quasiparticle weight $Z_\svek{k}$ in the $k_z=0$ plane for the $d_{xy}$ orbital.\cite{Note3}} 
\label{fig:siw_plane}
\end{minipage}
\end{figure*}

The results for the self-energy are displayed in the two top panels of Fig.~\ref{fig:siw_multiplot} for three selected $\svek{k}$-points and are compared to the momentum-independent DMFT self-energy.
We first discuss the self-energy via its low-frequency expansion: $\Sigma(\svek{k},i\nu)=\Re(\svek{k},i\nu\rightarrow0)+i\Im\Sigma(\svek{k},i\nu\rightarrow 0)+(1-1/Z_\svek{k})i\nu+\mathcal{O}(\nu^2)$.
From the local DMFT self-energy we extract\cite{Note3}
 a quasi-particle weight $Z^{\hbox{\tiny DMFT}}=0.49$ and a scattering rate $\gamma^{\hbox{\tiny DMFT}}\equiv-\Im\Sigma^{\hbox{\tiny DMFT}}(i\nu\rightarrow 0)=0.37\,$eV.
The imaginary parts of the AbinitioD$\Gamma$A Matsubara self-energy (see \fref{fig:siw_multiplot} top panel) suggest a slight enhancement
of the quasi-particle weight $Z_\svek{k}$ (smaller slope at low energy) for all momenta and orbital components.
Interestingly, we find for the quasi-particle weight $Z_\svek{k}$ an extremely weak momentum-dependence. 
Indeed $Z_\svek{k}$ varies by less than 2\% within the Brillouin zone. This is also illustrated in Fig.~\ref{fig:siw_plane}(d) which displays $Z_\svek{k}$ of the $d_{xy}$ Wannier orbital in the $k_z=0$ plane.
The corresponding dependence of $\gamma_\svek{k}$ is displayed in panel (c) of Fig.~\ref{fig:siw_plane}. Also here, we see only a small momentum differentiation of at most 10\%.

The momentum-dependence of the D$\Gamma$A self-energy in general further allows for an orbital differentiation of correlation effects in this locally degenerate system.\cite{NoteOffdiag}
For $Z_\svek{k}$ and $\gamma_\svek{k}$ that are both obtained from the {\it imaginary} part of the Matsubara self-energy, only a small difference between (at this $\svek{k}$) non-equivalent orbital components develops (see top panel in Fig.~\ref{fig:siw_multiplot}).

Much more sizable effects occur for both the momentum and the orbital dependence of the {\it real-part} of the self-energy at low energies. This can be inferred from the middle panel
of Fig.~\ref{fig:siw_multiplot} and Fig.~\ref{fig:siw_plane} (a) that displays $\Re\Sigma(\svek{k},i\nu_0)$ at the lowest Matsubara frequency,  again for the $d_{xy}$ orbital in the $k_z=0$ plane.
We witness a momentum-differentiation of 0.2eV or more---a quite notable effect beyond DMFT.
We note that, contrary to $Z_\svek{k}$ and $\gamma_\svek{k}$, the momentum-dependence of $\Re\Sigma(\svek{k},i\nu_0)$ in  \fref{fig:siw_plane} (a)  does not mirror the shape of the Fermi surface in \fref{fig2} (right).  This will in particular influence transport properties that probe states in close proximity to the Fermi surface.

At low energies, we also find a pronounced orbital-dependence in $\Re\Sigma(\svek{k},i\nu)$: At the $X$-point the real-part of the low-frequency self-energy is larger by about 0.1eV
for the (at this $k$ point) degenerate $d_{xy}$, $d_{yz}$ orbitals than for the $d_{xz}$ component. At the $M$ point the $d_{xy}$ component is larger than the $d_{xz}$, $d_{yz}$ doublet. 

Combining the influence of the orbital- and momentum dependent self-energy,
we hence find systematically larger shifts $\Re\Sigma(\svek{k},i\nu=0)$ for excitations with higher initial (DFT) energy. 
Seen relatively, this means that unoccupied states are pushed upwards
and occupied states downwards, resulting in a widening of the overall band-width.
This was previously evidenced using perturbative techniques.\cite{Miyake13,jmt_sces14,Tomczak14}

At high energies, the self-energy becomes again independent of orbital and momentum to recover the value of the Hartree term. \cite{Note5}

We now use the maximum entropy method \cite{Jarrell1996,Sandvik98b} to analytically continue the AbinitioD$\Gamma$A Green's function to real frequency spectra.
Let us note that, in our AbinitioD$\Gamma$A calculations we do not update the chemical potential. However, from the D$\Gamma$A Green's function we find a particle number of 1.062, which is very close to the target occupation of 1.

In the lowest panel of Fig.~\ref{fig:siw_multiplot} we compare our results to conventional DMFT for selected k-points.
From the above discussion it is clear that the AbinitioD$\Gamma$A self-energy will cause quantitative differences in the many-body spectra, while the overall shape will be qualitatively similar
to our and previous DMFT results. 
As evidenced above, the inclusion of non-local fluctuations decreases the degree of electronic correlations: 
Both a larger $Z$ and the shifts induced by $\Re\Sigma$ slightly increase the interacting band-width with respect to DMFT.
Indeed we see in our spectra signatures of reduced correlations: Hubbard bands are less pronounced and quasi-particle peaks move away from the Fermi level, although in the current case
these effects are small. This is congruent with previous dynamical cluster approximation (DCA) calculations that included short-ranged non-local fluctuations.\cite{Lee12}
Let us also note that recently it was indeed found experimentally,\cite{2016arXiv160206909B} that the lower Hubbard band in SrVO$_3$ is intrinsically somewhat less pronounced than previously thought,
with a substantial part of spectral weight actually originating from oxygen vacancies.

The very weak momentum dependence of the quasi-particle dynamics
and electronic lifetimes does not come as a surprise. Indeed the local nature of $Z$ was previously established
in a D$\Gamma$A study of the 3D Hubbard model\cite{Schaefer2015}, and, using perturbative techniques, in metallic oxides\cite{Tomczak14} and the iron pnictides and chalcogenides\cite{jmt_pnict,jmt_sces14}. 
On the other hand, these studies found a largely momentum-dependent static contribution $\Re\Sigma(\svek{k},\nu=0)$ to the self-energy.
Going beyond model studies and perturbative methods, we here confirm that $\Re\Sigma(\svek{k},\nu=0)$ indeed contains non-negligible momentum-dependent 
correlations beyond DMFT even for only purely local interactions.
Still, in the current study, momentum-dependent effects are small enough to only lead to quantitative changes.
There are three main reasons for the preponderance of local self-energy effects:
(1) SrVO$_3$ is not in close proximity to a spin-ordered phase or any other second order phase transitions. Therefore, non-local spin- or charge-fluctuations were not expected
to be particularly strong. (2) SrVO$_3$ is a cubic, i.e.\ fairly isotropic system. Non-local correlation effects are generally more pronounced in anisotropic or lower dimensional systems.
Therefore, we can speculate that non-local self-energies will become more prevalent in ultra-thin films of SrVO$_3$\cite{PhysRevLett.104.147601,PhysRevLett.114.246401}.
(3) The $GW$ approach in fact yields a much larger static $\svek{k}$-dependence $\Re\Sigma(\svek{k},\nu=0)$.\cite{Miyake13,Tomczak14}
This is however an effect of the non-locality of the {\it interaction} which yields a largely momentum-dependent screened exchange contribution to the self-energy. \cite{Note7}
While non-local interactions are included in the AbinitioD$\Gamma$A formalism (see Sec.~\ref{Sec:method}), we here performed calculations with a local interaction only, and are thus missing
this effect.

\section{Conclusion and outlook}
\label{Sec:outlook}

In conclusion we have derived, implemented, and applied a new first principles technique for correlated materials: the AbinitioD$\Gamma$A approach.
The method is a diagrammatic extension of the successful DMFT approximation and treats electronic correlation effects on all time and length scales.
Since it includes the self-energy diagrams of DMFT, the {\it GW} approach and non-local correlations beyond both, we believe AbinitioD$\Gamma$A to
set a new standard in realistic many-body calculations. We first applied the new methodology to the transition metal oxide SrVO$_3$ in a one-shot setup and neglected the influence of frequency dependent and non-local interactions, $U(\omega)$ and $V^q$, respectively.
Consequently the  plasmonic physics recently reported in $GW$+DMFT\cite{Tomczak2012,Tomczak14,Boehnke2016} is not included.
Here we focused on non-local correlation effects beyond DFT+DMFT that arise from a purely local Hubbard-like interaction such as non-local spin fluctuations.

We find that while the quasi-particle weight $Z$ is essentially local, there is a notable momentum- and orbital-dependence in the real part of the
self-energy. We hence conclude that non-local correlations can be important even in fairly isotropic systems in three dimensions, in the absence of any fluctuations associated with a nearby ordered phase,
and can occur even for purely local (Hubbard \& Hund) interactions.
These findings herald the need for advancing state-of-the-art methodologies for the many-body problem. In this vein, AbinitioD$\Gamma$A presents a very promising route toward the quantitative simulation of materials.
In future studies the approach can be applied to systems in which non-local fluctuations play a greater role, such as compounds in proximity to second order phase transitions or lower dimensional systems. For such materials non-local correlations beyond DMFT are a journey into the unknown.

{\em Acknowledgments.} 
We thank J. Kaufmann,  G. Rohringer, T. Sch\"afer and A. Toschi for useful discussions, as well as A. Sandvik for making available his maximum entropy program.   This work has been supported  by European Research Council under the European Union's Seventh
Framework Program (FP/2007-2013)/ERC through grant agreement n.\ 306447;
AG also thanks the Doctoral School W1243 Solids4Fun (Building Solids for Function) of the Austrian Science Fund (FWF). 
Calculations have been done on the Vienna Scientific Cluster~(VSC).

{\em Note added.} In the course of finalizing this work, we became aware of the independent development of a related
{\em ab initio} vertex approach by Nomura {\em et al.} \cite{Nomura2016} based on another diagrammatic DMFT extension, the triply-irreducible local expansion akin to D$\Gamma$A.


\begin{thebibliography}{91}
\bibitem{NotebeyondGW} 
Beyond $GW$, which only includes the particle-hole channel, also the diagrams of the particle-hole transversal channel are included. Moreover, in $GW$ only the first $V^q$ term of  Fig.\ \ref{Fig:flow} (b) is included, not the crossed  $V^{k-k'}$ which describes spin fluctuations. At least for the local $U$ part of the less correlated orbitals
this crossed term can be included  with ease, as it is independent of ${k-k'}$. It corresponds to having a different vertex for spin and charge contributions (see below). Such a combined scheme where the local vertex is calculated for some correlated orbitals and $U+V^q$ is taken for the less correlated orbitals is best combined with a $GW$ treatment to avoid double counting self-energies.
\bibitem{NoteVSVO}
Our constrained RPA estimate yields the following interactions for the $t_{2g}$ maximally localized Wannier setup: local Hubbard interaction $U=3.5\,$eV, nearest neighbor interaction $V_{NN}=0.8\,$eV, next-nearest neighbor interaction $V_{NNN}=0.5\,$eV. Note that this constrained RPA value for $U$ taken at $\omega=0$ is smaller than the constrained DFT  value employed in our calculation, which can be understood from the fact that  $U(\omega$) increases with $\omega$ so that a larger static $U$ is necessary to arrive at a similar level of correlation.
\bibitem{NoteOffdiag}
In particular, away from high-symmetry points, the lifting of degeneracy also allows for orbital-offdiagonal components in the self-energy. We however find these to be very small in the current system, which is why we limit the discussion to the diagonal components.
\end{thebibliography}

%

\end{document}